\documentclass[twocolumn]{aastex631}

\usepackage{amsmath}
\usepackage{wrapfig}
\usepackage{xcolor}

\newcommand{\taud}{\tau_{\rm d}}
\newcommand{\Nphi}{N_\phi}

\begin{document}

\title{Redeveloping a CLEAN Deconvolution Algorithm for Scatter-Broadened Radio Pulsar Signals}

\author[0000-0002-0883-0688]{Olivia Young}
\affiliation{School of Physics and Astronomy, Rochester Institute of Technology, Rochester, NY 14623, USA
}
\affiliation{Laboratory for Multiwavelength Astrophysics, Rochester Institute of Technology, Rochester, NY
14623, USA}

\author[0000-0003-0721-651X]{Michael T. Lam}
\affiliation{SETI Institute, 339 N Bernardo Ave Suite 200, Mountain View, CA 94043, USA}
\affiliation{School of Physics and Astronomy, Rochester Institute of Technology, Rochester, NY 14623, USA
}
\affiliation{Laboratory for Multiwavelength Astrophysics, Rochester Institute of Technology, Rochester, NY
14623, USA}



\begin{abstract}

Broadband radio waves emitted from pulsars are distorted and delayed as they propagate toward the Earth due to interactions with the free electrons that compose the interstellar medium, with lower radio frequencies being more impacted than higher frequencies. Multipath propagation in the interstellar medium results in both later times of arrival for the lower frequencies and causes the observed pulse to arrive with a broadened tail described via the pulse broadening function. We employ the CLEAN deconvolution technique to recover both the intrinsic pulse shape and pulse broadening function. This work expands upon previous descriptions of CLEAN deconvolution used in pulse broadening analyses by parameterizing the efficacy on simulated data and developing a suite of tests to establish which of a set of figures of merit lead to an automatic and consistent determination of the scattering timescale and its uncertainty. We compare our algorithm to simulations performed on cyclic spectroscopy estimates of the scattering timescale. We test our improved algorithm on the highly scattered millisecond pulsar J1903+0327, showing the scattering timescale to change over years, consistent with estimates of the refractive timescale of the pulsar.

\end{abstract}


\section{Introduction}

%Radio pulsars provide unique probes of the ionized interstellar medium (ISM), allowing us to gain insight into its structure and variability by modeling the effects on the delays and distortions on the emitted radio pulses as observed at the Earth \todo{citations}. While delays due to dispersion are routinely modeled in pulsar timing experiments \todo{citations}, distortions due to multipath propagation are not. The inverse problem is difficult in that the intrinsic pulse shape is unknown and the underlying geometry and spectrum of the turbulent medium is both unknown and the path traced through the medium can cause variations in the observed pulse broadening function (PBF) over time \todo{citations}. Not only can separating these two effects yield important insights into the nature of the ionized ISM, we can also properly mitigate the impact on the pulse profiles used in precision pulsar timing experiments such as those used in low-frequency gravitational wave detectors \todo{citations}.

Radio pulsars provide unique probes of the ionized interstellar medium (ISM) and allow us to gain insight into its structure and variability by modeling the effects of the delays and distortions on the emitted radio pulses as observed at the Earth \citep{pulsar_handbook}. While delays due to dispersion are routinely modeled in pulsar timing experiments \citep[e.g., ][]{vname}, distortions due to multipath propagation are not and it can be difficult to do so \citep{sc2017}. Determining the distortion level is difficult due to both the intrinsic pulse shape and the underlying geometry and spectrum of the turbulent medium being unknown \citep{cpl1986,cr1998}, and the time and path-dependent variations in the observed pulse broadening function \citep[PBF; ][]{Williamson}. Not only can separating these effects yield important insights into the nature of the ionized ISM but also provide proper pulse profile impact mitigation for pulsars used in precision timing experiments such as low-frequency gravitational wave detectors \citep{Stinebring2013}.

CLEAN deconvolution, originally developed for radio interferometric imaging \citep{Hogbom_1974}, was applied to radio pulses in \citet{Bhat_2003} to recover both the pulse broadening (scattering) timescale $\taud$ and the intrinsic shape simultaneously via the use of an assumed PBF. Unlike in synthesis imaging where the positions of the array elements are known while the sky brightness distribution is not, neither the analogous PBF nor intrinsic pulse shape, respectively, are known. \citet{Bhat_2003} introduced figures of merit to iteratively test trial values of $\taud$ under an assumed PBF, demonstrating variation in the rebuilt intrinsic pulses for PSR J1852+0031 for different PBFs and application to several other pulsars.

We expand upon the CLEAN deconvolution algorithm presented in \cite{Bhat_2003} to prepare for {\it automated} deployment on data sets of significantly more pulsars. In this work, we primarily focus on the broadening effects of the ISM and recovering $\taud$ with the intention of applying the algorithm to the multi-frequency profiles of pulsars distributed throughout the galaxy to understand both the bulk properties of the turbulence in the ISM and specific unique lines of sight. Understanding these properties inform priors on pulsar timing arrays and other high-precision pulsar timing experiments in which scattering biases estimates of the arrival times \citep{Lentati}. This work is the first of several papers on robust method development and deployment on real data from a larger selection of pulsar observations.

In \S\ref{sec: CLEAN}, we describe the CLEAN deconvolution method as presented and expanded upon the work in \citet{Bhat_2003}. In \S\ref{sec:sim_data}, we perform systematic tests on simulated data, demonstrating the level of recall in the input $\taud$ values and quantifying our uncertainties in the estimates. We also compare our results with the cyclic spectroscopy (CS) deconvolution technique and discuss the tradeoff of limitations in our method with the extensive computational complexity of the CS method. Finally, we apply our method to PSR J1903+0327 in \S\ref{sec:J1903} and discuss our future directions in \S\ref{sec:future}.

\section{The CLEAN Deconvolution Algorithm}

\label{sec: CLEAN}

%In this section, we will discuss CLEAN deconvolution as presented in \cite{Bhat_2003}.  Although the one-dimensional nature of pulsar profiles is simplistic compared to images from interferometers comprised of tens to hundreds of telescopes, there is one drastic, and important, difference between the two approaches: in traditional CLEAN applications, the instrumental point source response function is known, whereas the analogous function in this work, the pulse broadening function, is not. So while we still perform an iterative deconvolution in this work, we must assume the pulse broadening function using proposed models of the ISM composition. 

CLEAN deconvolution for radio pulsars exploits the one-dimensional nature of pulsar profiles and differs from traditional CLEAN approaches where the instrumental response function is known. The analogous function in this work, the PBF, must be assumed from {\it a priori} models. \citet{Bhat_2003} developed a method that can both determine the pulse broadening timescale $\taud$ and recover the intrinsic pulse from observational pulsar profile data via the employment of a CLEAN deconvolution algorithm and figures of merit (FOMs). CLEAN can be applied using different models of the PBF of the ISM, making it a broadly encompassing method. In this work, we assumed the PBF for the commonly-used thin-screen approximation for the ISM's geometry.

%Although the authors assume an intrinsic pulse shape that minimizes symmetry and a 
%\todo{I moved this paragraph up, I'm not sure where else it would belong? \citet{Bhat_2003} discuss the components an observed pulsar pulse is comprised of, introduce the methodology of their algorithm, and the five figures of merit used to determine the correct broadening timescale from a set of test values and apply their method to both real and simulated data.}
%In \citet{Bhat_2003} the CLEAN algorithm, along with five FOM used to determine the correct broadening timescale from a set of test values, were described. 
%In this section, we first describe the CLEAN algorithm along with the FOMs used in \citet{Bhat_2003} and developed here to determine the correct broadening timescale from a set of test values.

\citet{Bhat_2003} described the CLEAN algorithm for use in the deconvolution of radio pulsar pulses, along with the development of five FOMs used to determine the correct broadening timescale from a set of test values. In this section, we discuss the algorithm both as originally described and how the algorithm has been redeveloped for this work.

\subsection{Modeling the Observed Pulse Profile}

We assumed the observed pulse $y(t)$ to result from the convolution of the intrinsic pulse $x(t)$, the PBF $g(t)$, and the instrumental response function $r(t)$, given by
\begin{equation}
   y(t) = x(t) \otimes g(t) \otimes r(t).
\end{equation}
We simulated our intrinsic pulse $x(t)$ as a normalized, single-peaked Gaussian shape, which minimizes the asymmetry of the rebuilt pulse and provides a baseline comparison against the use of the FOM $\Gamma$ discussed later in Section~\ref{sec:shapeFOMs}. 

The PBF for the ISM is commonly modeled as a thin screen \citep{cr1998} for simplicity. The thin-screen approximation simplifies calculations, separating out the physical turbulent processes from the geometry of the intervening gas, and in the case of the PBF, simplifies the form as well; the thin-screen model works reasonably well for lines of sight with a single overdense region. We used this model in our work, given by
\begin{equation}
    g(t|\taud) = \frac{1}{\taud} \exp \left(-\frac{t}{\taud}\right) U(t),
\label{eq:PBF}
\end{equation}
where $U(t)$ is the Heaviside step function.

Lastly, the instrumental response function denoted as $r(t)$ determines the resolution of the observed data. We assumed a delta function\footnote{For clarity, we use the digital signal processing definition of the unit-height sample function being $\delta(t) = 1$ if $t=0$, otherwise 0, which allows us to multiply by a constant as in Eq~\ref{eq:CC}.} as an approximation for the instrumental response function with a width of one phase bin.

\subsection{CLEAN Deconvolution}\label{sec:form}

CLEAN iteratively subtracts replicated components from an observed pulse until the residual structure falls below the root mean squared (rms) of the off-pulse noise. As we do not know the value of $\taud$ {\it a priori}, this iterative subtraction process is repeated for a range of test $\taud$ values, with the assumed correct $\taud$ chosen using FOMs. For the purposes of the algorithm, we treat $\taud$ to be measured in time-bin resolution units as measured across the folded pulse's phase with $N_\phi$ total bins. We step through our CLEAN deconvolution process below.

%CLEAN is an iterative process, where small replicas of an observed pulse, called CLEAN components, are subtracted off the observed pulse until the residual falls below the root mean squared of the off pulse noise. This process is done for a range of different test $\taud$ values, with the correct value being chosen using a set of figures of merit.

%We will now step through the CLEAN deconvolution method as described in \citep{Bhat_2003} for simulated data. 

\begin{enumerate}

%In \cite{Bhat_2003}, the intrinsic pulse is a symmetric pulse, which in this work will we will assume to be a normalized, single-peak Gaussian pulse, denoted as $x(t)$. We acknowledge that this assumption is based on a simplified understanding of pulsar emission, and will not ensure a direct translation of the algorithm's performance from simulated to observational data.   
\item \textbf{CLEAN Component Creation:} 
%For each CLEAN iteration, a CLEAN component (CC) is subtracted off the main pulse of the input profile and used to rebuild the intrinsic pulses after the termination of the CLEAN iterations.  
We first identify the location of the maximum of the deconstructed pulse after the $i$-th iteration, $t_i \equiv \operatorname*{argmax}\left[y_i(t)\right]$; our first iteration begins with the originally observed pulse $y_0(t)$.  Each CLEAN component (CC) $y_c(t|t_i)$ starts with a delta function $\delta(t-t_i)$ at the location of the maximum of the observed pulse, $\max\left[y_i(t)\right]$ multiplied by the loop gain value $\gamma$, i.e., 
\begin{equation}
y_c(t|t_i) = \gamma \left\{ \max\left[y_i(t)\right] \right\} \delta(t-t_i) \equiv C_i \delta(t-t_i).
\label{eq:CC}
\end{equation}
Smaller loop gains result in a greater number of iterations before the stopping criterion is met but allow for finer intrinsic features to be resolved \citep{Hogbom_1974}; in this work, we used $\gamma = 0.05$.

\item \textbf{Iterative Subtraction off the Main Pulse:} After we construct $y_c(t|t_i)$, we convolve the CC with the instrumental response function $r(t)$ and the PBF with a given test $\taud$, and then subtract this shape from the $i$-th iteration pulse. The change in the profile at each iteration is described as
\begin{equation}
   \Delta y_i(t) = y_i(t) - \left\{ y_c(t|t_i) \otimes \left[g(t|\taud) \otimes r(t)\right] \right\}
\end{equation}
with $y(t_i)$ as the input pulse profile to the $i$-th iteration. The CCs are then iteratively subtracted off the main pulse, with the resulting subtracted profile becoming the pulse profile for the next CLEAN iteration so that
\begin{equation}
y_{i+1}(t) = \Delta y_i(t).
\end{equation}

\item \textbf{Termination of CLEAN Algorithm:} The CLEAN algorithm is terminated when the maximum of the input pulse profile falls below the rms of the off-pulse noise, i.e., $\max\left[y_i(t)\right] \le \sigma_{\rm off}$.

%, with the CCs used to rebuild the intrinsic pulse. 
\end{enumerate}

The CLEAN algorithm above will provide the list of CCs along with the residual noise. The CCs can be used to reconstruct the intrinsic pulse shape, but for the purposes of this work, our final goal was to determine $\taud$. The algorithm can run with any input value of $\taud$, therefore our iterative method is repeated with different trial $\taud$, from which we derived FOMs based on the reconstructed intrinsic pulse shape and the residual noise that resulted from each trial $\taud$.

% This might now be redundant, will have to see
%\textbf{Determination of $\taud$ using the Figures of Merit:} Six figures of merit are used to determine the correct $\taud$, five outlined in \citet{Bhat_2003} and one developed for this work. Here, we will briefly describe each of the FOM, referring to Fig. \ref{fig: all_FOM} for visual reference. 

\begin{figure}[h!]
 
\centering 

       \includegraphics[width = \linewidth]{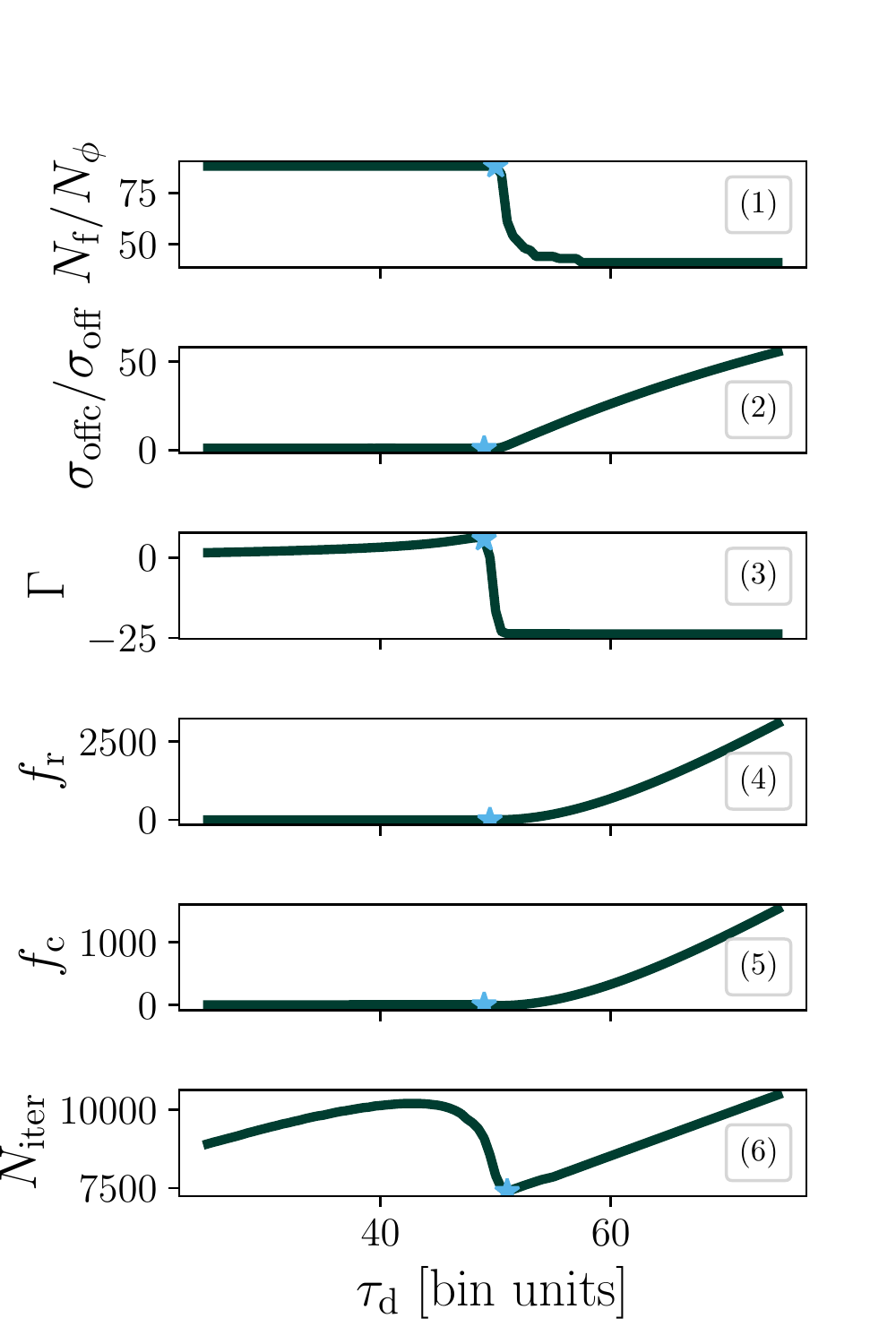}
       \caption{\label{fig: all_FOM} Summary of FOMs used in this work, for a pulse with simulated full-width at half maximum of 100 phase bin units, input $\taud = 50$ phase bins, and $\mathrm{S/N} = 100$. We tested $\taud$ values ranging from 25 to 75 bin units with a step size of one. Panel 1 shows the number of data points within a 3$\sigma$ level of the noise FOM, panel 2 shows the root mean squared FOM, panel 3 shows the skewness FOM, panel 4 shows the positivity FOM, panel 5 shows the combined skewness and positivity FOM, and panel 6 shows the number of iterations FOM. }
\end{figure}

\subsection{Figures of Merit}\label{sec:form}

%positivity

We employed six FOMs as follows: a measure of positivity of the residual noise ($f_{\rm r}$), a measure of skewness of the recovered intrinsic pulse ($\Gamma$), a count of the on-pulse-region residual points below the off-pulse noise level ($N_{\rm f}$/$N_{\rm \phi}$), a measure of the ratio of the rms of the residual noise to the off-pulse noise rms ($\sigma_{\rm offc}$/$\sigma_{\rm off}$), a measure of the combined positivity and skewness measure ($f_{\rm c}$), and a count of the number of CLEAN components each test $\taud$ uses before the peak of the profile falls below the noise level ($N_{\rm iter}$). All except the last are described in \citet{Bhat_2003}. These six FOMs fall into three broad categories: figures based on the rebuilt intrinsic pulse, figures based on the residual noise after the CLEAN algorithm terminates, and a figure based on the number of CLEAN components generated before the algorithm terminates. We describe the FOMs grouped into these three categories in the sections below.

In Figure \ref{fig: all_FOM}, we see the ideal result of the use of six FOMs and the methods for determining the ``correct'' $\taud$. % FOMs where the shapes all align at the location of the correct $\taud$. 

\subsubsection{FOM Measuring the Shape of the Rebuilt Intrinsic Pulse}
\label{sec:shapeFOMs}

We examine the CC amplitudes $C_i$ and locations $t_i$ found during the CLEAN process (e.g., see Eq.~\ref{eq:CC}) to compute the $\Gamma$ FOM. In our simulations, we created intrinsic pulses that are symmetric Gaussians, and therefore the correct rebuilt pulse should always be a perfectly symmetric Gaussian if the correct $\taud$ is used. In reality, intrinsic pulses may not be perfectly symmetric, and we discuss these implications in \S\ref{sec:future}.

The $\Gamma$ of the rebuilt pulses is calculated for each test $\taud$ by computing the third standardized moment 
\begin{equation}
   \Gamma = \frac{\left<t^3\right>}{\left< t^2\right>^{3/2}},
\end{equation}
where $\left<t^n\right>$ is
\begin{equation}
   \left<t^n\right> = \frac{\sum_{i = 1}^{n_c} (t_i - \bar{t})^n C_i}{\sum_{i = 1}^{n_c} C_i}
\end{equation}
and $\bar{t}$ is
\begin{equation}
   \bar{t} = \frac{\sum_{i = 1}^{n_c} t_i C_i}{\sum_{i = 1}^{n_c} C_i}.
\end{equation}
The resulting $\Gamma$ is ideally represented by the example in panel 3 of Figure \ref{fig: all_FOM}, where the sharp fall-off point represents the general location of the correct $\taud$.

%We measure the skewness for each test $\taud$ we iterate through, with the resulting FOM ideally being represented as the example in panel 3 Figure \ref{fig: all_FOM}, where the sharp fall-off point represents the general location of the correct $\taud$.

\subsubsection{FOMs Based on the Residual Noise}

Three of our FOMs are built from measures of the residual noise after the completion of the CLEAN algorithm. We will also discuss a FOM that combines one of these FOMs (positivity) with the $\Gamma$ FOM discussed previously -- this is an important FOM as described in \citet{Bhat_2003}.

%There are three FOM that are built from different measures of the residual noise after the completion of the CLEAN algorithm: the positivity, number of data points within 3$\sigma$ of the noise, and the rms measure FOM. We will also discuss the combination of the skewness and positivity FOM in this section.

\begin{figure}[t!]
\centering 

       \includegraphics[width = 1.1\linewidth]{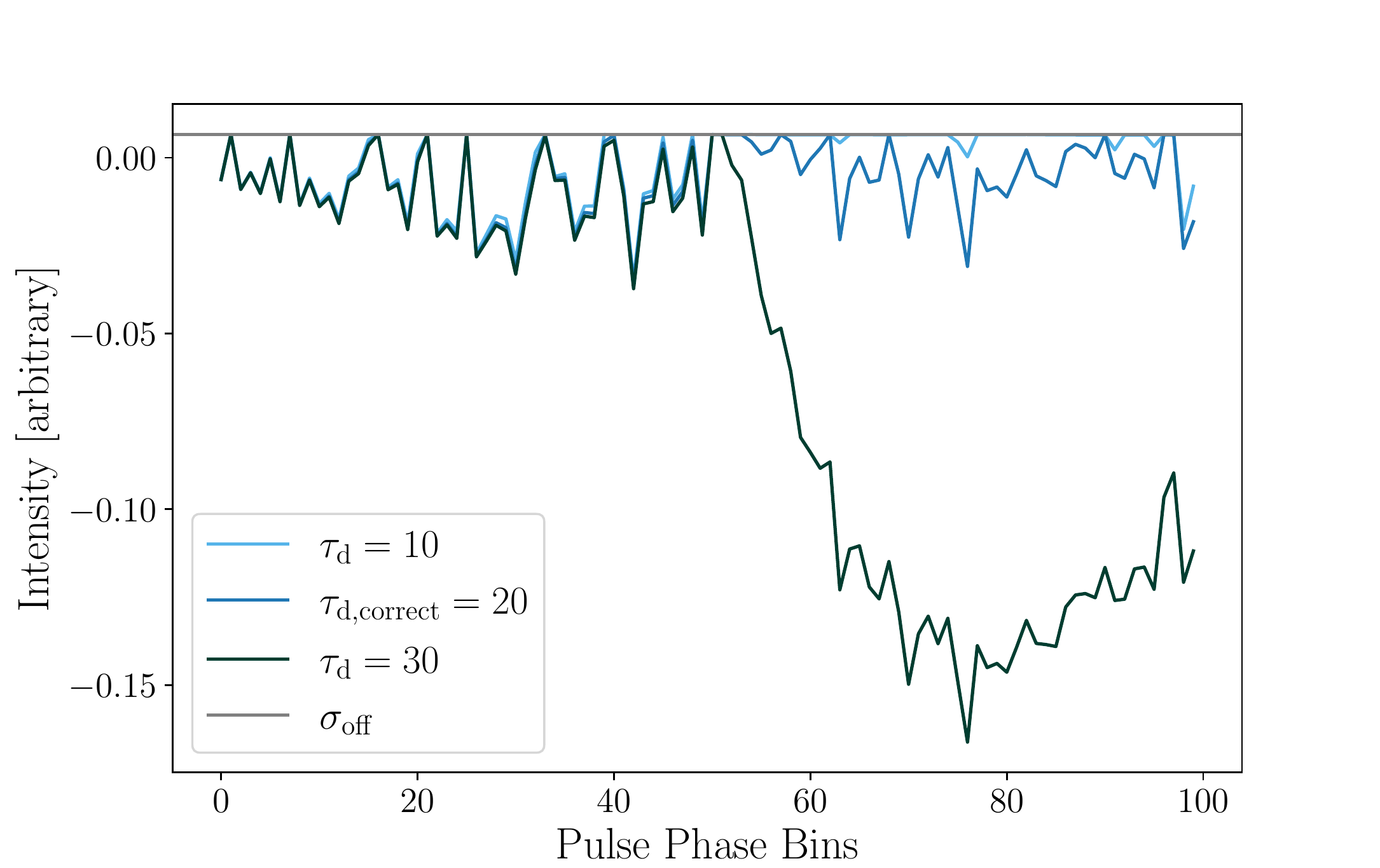}
       \caption{\label{fig:res_noise_compare} The residual noise left over after the CLEAN algorithm terminates for three test $\taud$ values, 10, 20, and 30 bins, where 20 is the simulated value.
        These time series are representative of the residuals used to calculate multiple FOMs. We can see the under- and over-subtraction for test $\taud$  values that are smaller than or larger than the true $\taud$, respectively.  }
\end{figure}

The residual noise is one of the end products of the CLEAN deconvolution process. A test $\taud$ that is larger than the correct value of $\taud$ results in a progressively larger over-subtraction as shown in Figure~\ref{fig:res_noise_compare}. If the test $\taud$ is smaller than the correct value, it results in an unremoved noise floor in the baseline, again see Figure~\ref{fig:res_noise_compare}.

We can first calculate the rms of the residual noise
\begin{equation}
\sigma_{\rm offc} = \frac{1}{\Nphi}\sum_{j = 1}^{\Nphi} [\Delta y_i (t_j|\taud) ]^2,
\end{equation}
in comparison to the rms of the off-pulse region, $\sigma_{\rm off}$, where this ratio $\sigma_{\rm offc}/\sigma_{\rm off}$ will grow whenever over- or under-subtraction is performed and should otherwise approach a value of 1 for the appropriate subtraction. This is roughly equivalent to the single metric used to automatically determine $\taud$ used in \citet{Tsai_2017} for multi-frequency data from 347 pulsars.

Beyond the rms, we can count the total number of residual noise points $N_{\rm f}$ within a certain threshold level (we chose $3\sigma_{\rm off}$) of the noise that satisfies the condition
\begin{equation}
   |y_i - y_\mathrm{off}| \leq 3\sigma_\mathrm{off}.
\end{equation}
As seen in Figure~\ref{fig:res_noise_compare}, for under-subtraction we expect all of the points to satisfy the condition and so the ratio $N_{\rm f}/\Nphi = 1$, but the ratio will drop as over-subtraction occurs.

%We will now look individually at each figure of merit that relies on the residual noise, as well as the combined skewness and positivity figure of merit.

Besides these two metrics which measure deviations away from the rms noise, we also wish to enforce non-negativity of the residual profile since we know pulsar signals must be above the baseline noise. A $f_{\rm r}$ FOM was defined\footnote{\citet{Bhat_2003} introduced a multiplicative weight of order unity but did not specify the value. Here we take that weight to be 1 and so ignore introducing it in the main text. They also include a Heaviside step function, $U_{\Delta y}$. As this only changes the overall normalization of our FOM in our simulation runs, we ignore this in our work.} by \citet{Bhat_2003} in terms of a sum over the $\Nphi$ bins of the residual noise, 
\begin{equation}
   f_{\rm r} = \frac{1}{\Nphi \sigma_\mathrm{off}^2} \sum_{j = 1}^{\Nphi} [\Delta y_i (t_j|\taud) ]^2 .
\end{equation}

If $\Delta y_i(t)$ is Gaussian white noise with rms equal to $\sigma_{\rm off}$, then as with the previous FOM, we would expect $f_{\rm r} \approx 1$, while over-subtraction would force the sum to increase well beyond 1. 
%We do not employ a Heaviside step function in this work due to our normalization of our FOM in our simulation runs.}

%in this
%\begin{equation}
%U_{\Delta y} \equiv
%\begin{cases}
%1, & $t > $\\
%0, & $t \leq $
%\end{cases}
%\end{equation}

%Here, $x$ is arbitrarily chosen to be $3/2$, and $m$ is a weight of order unity chosen to be 1 \cite{Bhat_2003}. The residual noise left over a the end of the iterative CLEAN process is denoted as $\Delta y$, $\theta$ is the root mean squared of the off-pulse noise, and $U$ is a Heaviside function used to account for the binning of the data. We see the expected shape of this FOM in panel 4 Figure \ref{fig: all_FOM}. 

\citet{Bhat_2003} defined the $f_{\rm c}$ FOM, equally weighting the rebuilt intrinsic pulse shape and the residual noise by
\begin{equation}
   f_{\rm c} = \frac{\Gamma + f_{\rm r}}{2},
\end{equation}
thus providing higher confidence in test $\taud$ values with favorable values of both skewness and positivity. The typical shape of this FOM is shown in panel 5 of Figure \ref{fig: all_FOM}.

%the two categories of FOM from the original paper, one based on the rebuilt intrinsic pulse and one based on the residual noise, the skewness and positivity metrics respectively. Therefore, test $\taud$ value with favorable values of both skewness and positivity will be weighted more highly in the end through the employment of this figure of merit. We see the expected shape of this FOM in panel 5 Figure \ref{fig: all_FOM}. 

%We use the rms of the off-pulse region of the observed pulse as the cut-off value for our CLEAN iterations, as well as comparing it against the rms of the residual noise after the CLEAN process has finished. 

\subsubsection{FOM Measuring the Number of Iterations Performed}

We developed this FOM to more directly measure the fit of the re-convolved CCs broadening tails to the broadening of the observed pulse. As the amplitude for re-convolved CCs with larger broadening tails is smaller than those with smaller broadening tails (due to the normalization in Eq.~\ref{eq:PBF}), we expect a general increase in the number of iterations needed to deconvolve the observed pulse. Similarly, when re-convolved CCs with smaller broadening tails are subtracted from a pulse with a larger true broadening tail, more iterations will be required. However, when the CCs are convolved with the correct value of $\taud$, neither under nor over-subtraction occurs, resulting in fewer iterations being needed. Therefore, we expect a dip in our FOM around the correct value of $\taud$. 

%The trial $\taud$ value affects both the amplitude and broadening of the CLEAN components, with lower values of $\taud$ needing fewer iterations needed for the pulse to fall below the noise level due to their larger amplitude and the number of iterations needed increasing with the value of $\taud$.

%As the trial $\taud$ value approaches the correct $\taud$, the slope of the CLEAN component will more closely align with the slope of the broadening tail of the pulse profile, thus requiring fewer CCs for the pulse to fall below the noise. 
%We expect a general increase in the number of iterations needed, with a dip around the correct value of $\taud$.
 %due to the balancing of the amplitude and slope effects. 

%However, as the value of test $\taud$ begins to get close to the correct value, the slope of the CLEAN component will begin to more closely match the slope of the broadening tail of the pulse profile, thus requiring fewer CCs in fewer locations for the pulse to fall below the noise. Therefore, we would expect a general increase in the number of iterations needed, with a dip around the correct value of $\taud$ due to the balancing of the amplitude and slope effects, and this is in fact what we see in the simulated data. We see the expected shape of this FOM in panel 6 Figure \ref{fig: all_FOM}.

\subsection{Automating the Choice of the Correct $\taud$ Value}

%\todo{somewhere in here, need to also discuss how the uncertainties are computed}

These FOMs were originally constructed to pinpoint the correct value of $\taud$ by eye. This approach is impractical for large data sets, so we automated this process. We found that the simple approach of computing the numerical third derivative of each function with respect to $\taud$ and finding the maximum has yielded good results, though the exact recall depends on both the value of $\taud$ and the pulse S/N. More complicated algorithms will be employed in future works but the systematic error introduced by this choice is small in comparison to other noise sources, as shown next, so we opted to use it.

%This method performs fairly well on our simulated data. The one exception to this approach is the number of iterations FOM were we are looking for a sharp dip in the FOM around the location of the correct $\taud$ instead of the location where the slope changes the most.

%While the goal of using these figures of merit is to have distinctly recognizable features to pinpoint the correct value of $\taud$ using the human eye, it is impractical to review thousands of these FOMs by hand. Therefore, a method of automation must be decided upon. In this work we decide to take the third derivative of the function with respect to $\taud$, to determine the location of the correct $\taud$. As all of the figures of merit from the original paper rely on a visual, drastic slope change to determine the correct $\taud$, the third derivative will return the $\taud$ where the slope has the greatest rate of change. This method performs fairly well on our simulated data. The one exception to this approach is the number of iterations FOM. For this, as we are looking for a sharp dip in the FOM around the location of the correct $\taud$ instead of the location where the slope changes the most, we choose to use the approach of finding the point where the difference between the point of interest and those before and after it is the greatest, thus signifying the location of the most drastic dip in the shape of the FOM.

\section{Automated Algorithm Performance}
\label{sec:sim_data}

%We have reproduced a CLEAN deconvolution algorithm written in Python to accurately describe the amount of broadening due to the ISM, $\taud$, in an observed pulsar profile following the methods described in \citet{Bhat_2003}. 

In this section, we will discuss the performance of our automated CLEAN method.
%and comparison between CLEAN and cyclic spectroscopy as outlined in \citep{Dolch_2021}, as well as results on observational data for pulsar J1903+0327. 
An in-depth description of our redeveloped CLEAN algorithm in Python, as well as notes on how to use the open source versions available on \url{https://zenodo.org/badge/latestdoi/524167339}, can be found in \citet{Young_2022}.

We wished to robustly quantify the ``correctness'' of our $\taud$ estimates in simulated data so that we could automatically assign uncertainties to our estimates on real data. To that end, we simulated multiple data sets with different input parameters to determine how these will affect the recall. Ideally, as in \citet{Dolch_2021} for the cyclic spectroscopy (CS) algorithm, only the S/N and $\taud$ of a profile should affect the recall accuracy of our CLEAN deconvolution, though we tested several other parameters as well.

To quantify the algorithm's performance, we computed a measure of the fractional average error bar. Within this work, the values returned for each FOMs were given the same weight when calculating our error bars, which is defined as the fraction of the returned $\taud$ to the correct injected $\taud$. Our fractional average error bars are defined as
\begin{equation}
    \epsilon_{\rm ave} = 1 - \frac{1}{N_{\rm runs}} \sum_{i=1}^{N_{\rm runs}}\left(\frac{\epsilon_{i}}{\taud N_{\rm FOM} }\right)
\end{equation}
where  $N_{\rm runs}$ is the number of simulations for a given data set, $ N_{\rm FOM} = 6$ is the number of FOMs used, and $\epsilon_{i}$, for readability, is defined as an unweighted sum of the $\taud$ values returned by our FOMs for each run ($i = 1 \dots N_{\rm runs}$),
\begin{equation}
    \epsilon_{i} = {\tau_{\frac{N_{\rm f}}{{N_\phi}}} + \tau_{\frac{\sigma_{\rm offc}}{\sigma_{\rm off}}} + \tau_{\Gamma} + \tau_{ f_{\rm r}} + \tau_{f_{\rm c}} + \tau_{N_{\rm iter}}}.
\end{equation}

\subsection{Testing the Impact of S/N and $\taud$ on Recall}
% and a Comparison to Cyclic Spectroscopy}
%\label{sec:CS}

We first tested how CLEAN performs based on different injected pulse S/N and $\taud$ combinations. We simulated data sets using several of the characteristics of PSR~B1937+21 (the first-known millisecond pulsar and a known scattered source) as follows; these parameters are also shown in Table \ref{table:sim_1}. 
PSR~B1937+21 has a spin period of 1.557~ms and a full-width at half maximum (FWHM) of 38.2~$\mu$s \citep{Manchester+2013}. To reduce computing time, we used different numbers of phase bins depending on the injected $\taud$ value as shown in Table~\ref{table:sim_1}; we show in the next subsection that there is minimal impact in the recovery of $\taud$ depending on the phase resolution of the pulses so long as the scattering tails are resolved. For each of our runs, we tested across 100 equally-spaced $\taud$ steps between 0.5 and 1.5 times the injected $\tau_{\rm d, correct}$. For each S/N-$\taud$ pair, we simulated and ran CLEAN on 60 pulse shapes.

%These results are from simulated data sets, using the characteristics of PSR B1937+21. We incorporate all components used in \citep{Dolch_2021} into our parameterization data set, except for the number of bins. In \citep{Dolch_2021}, 2048 bins are used for all runs. As shown in our small-scale parameterizations, we have determined that our recall doesn't change significantly when using 512, 1024, or 2048 bins. To conserve computing resources, we have employed a staggered binning grid to both reduce the run time and preserve the resolution needed to resolve each set of $\taud$ values.

%Cyclic spectroscopy is also used to recover the pulse broadening function of radio pulsars. In this work, we compare our results to those from \cite{Dolch_2021}, where the authors parameterize the performance of cyclic spectroscopy using a range of $\taud$ and S/N values. 

We choose our S/N-$\taud$ values in the same style as \citet{Dolch_2021} to more directly compare our CLEAN deconvolution method with the CS algorithm. While CLEAN works on an averaged pulse profile, CS uses raw voltage data prior to folding to recover the full impulse response function, making the latter more computationally intensive (though assuming no specific PBF). These methods are therefore difficult to directly compare, but we can expect to see improved performance for both methods as either S/N or $\taud$ increase.

%CLEAN uses a pulse profile, or a folded average of the pulse across an observation, that is iteratively deconvolved and then chooses $\taud$ based on the shapes of six figures of merit. Each FOM is built upon measurements of the reconstructed pulse, the residual noise, or in this work the number of CLEAN iterations. Cyclic spectroscopy, on the other hand, uses strings of individual pulses, or unfolded pulsar data, and raw voltage data, making this method computationally intensive. Each pulse is evaluated separately, with the confidence in the chosen $\taud$ becoming greater with every cycle. 

\begin{figure}[h!]
\centering 

       \includegraphics[width = 1.1\linewidth]{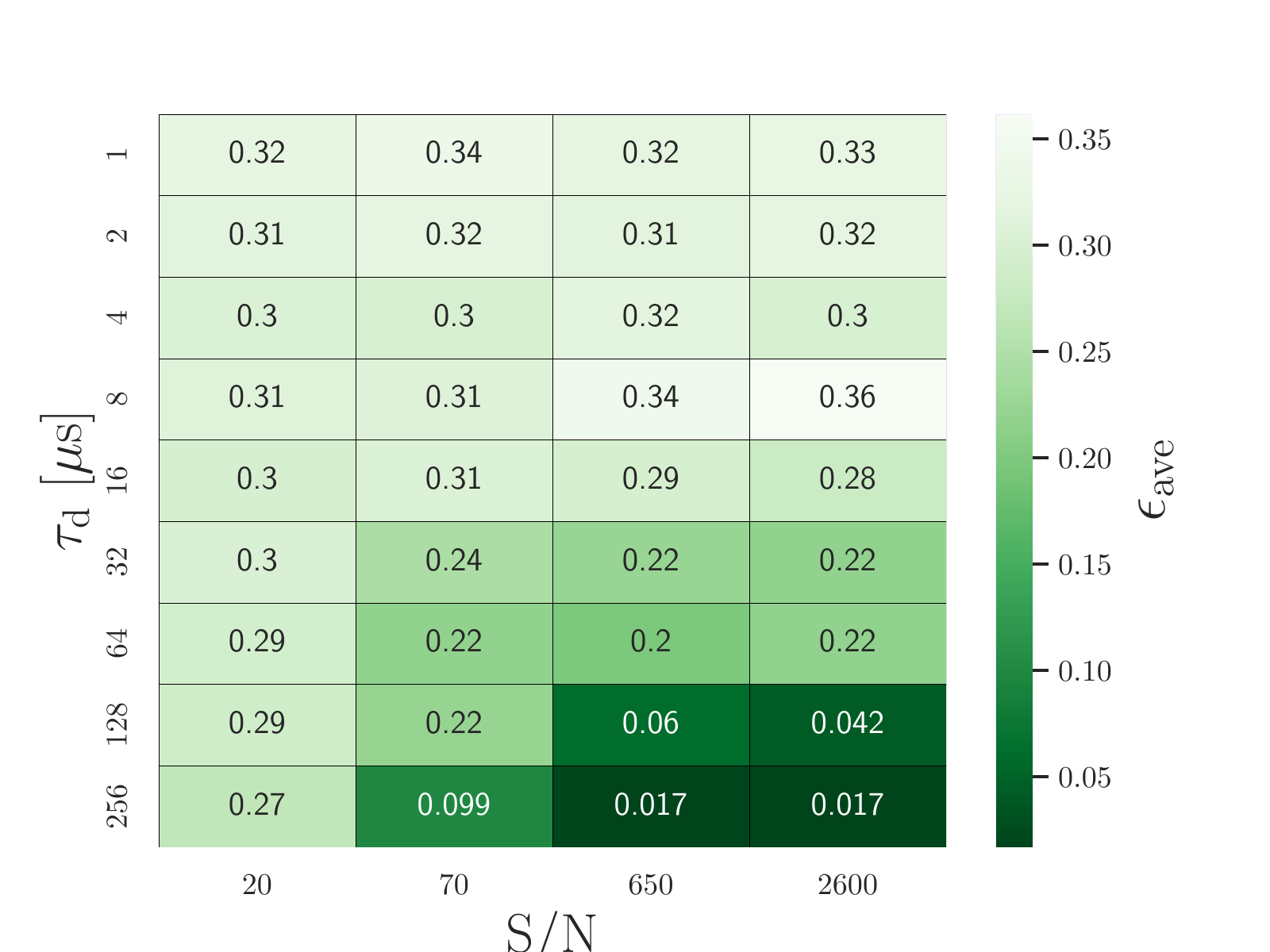}
       \caption{\label{fig:error_bar} Average recall error bars for CLEAN deconvolution in  the same style as Figure 5 from \citet{Dolch_2021}. This plot gives an encompassing overview of the performance of the CLEAN algorithm by returning the average size of the error bars with each S/N and $\taud$ pair. As smaller error bars indicate better performance, CLEAN performs better on simulated data with larger values of both S/N and $\taud$. }
\end{figure}

Indeed, after running our simulations, we see this expected behavior in Figure \ref{fig:error_bar}, where darker colors indicate better recovery of $\taud$, which matches what is seen in \citet{Dolch_2021} for CS. The numerical values shown in Figure~\ref{fig:error_bar} are in terms of the percentage of the correct $\taud$, given by $\epsilon_{\rm ave}$.

%In Figure \ref{fig:error_bar}, we see the average recall error bars of our runs. Our error bars are in terms of the percentage of the correct $\taud$. Each error bar returned is the averaged value returned by each of the FOM for each run averaged across 60 runs, or the average of 360 independently determined $\taud$ values. 

%for cyclic spectroscopy in Figure 5 in \cite{Dolch_2021} and in Figures \ref{fig:error_bar} and \ref{fig:FOM} for CLEAN, with darker colors indicating better recovery. These results are from simulated data sets, using the characteristics of PSR B1937+21. We incorporate all components used in \citep{Dolch_2021} into our parameterization data set, except for the number of bins. In \citep{Dolch_2021}, 2048 bins are used for all runs. As shown in our small-scale parameterizations, we have determined that our recall doesn't change significantly when using 512, 1024, or 2048 bins. To conserve computing resources, we have employed a staggered binning grid to both reduce the run time and preserve the resolution needed to resolve each set of $\taud$ values. These parameters are shown in Table \ref{table:my_parm}.

\begin{deluxetable}{lc}
\centering 
\tablecolumns{2}
\tablecaption{Automated Algorithm Simulation Parameters}
\tablehead{
\colhead{Parameter} & \colhead{Value}
\label{table:sim_1}
}
\startdata
  Spin period & 1.557 ms \\
 Pulse FWHM & 38.2~$\mu$s \\
  $\Nphi$ for $\taud = 1,2,4~\mu \mathrm{s}$ & 2048 \\ 
 $\Nphi$ for $\taud = 8,16, 32~\mu \mathrm{s}$ & 1024 \\ 
 $\Nphi$ for $\taud = 64, 128, 256~\mu \mathrm{s}$ & 512 \\ 
  Test $\taud$ range & (0.5$\tau_{\rm d, correct}$ -- 1.5$\tau_{ \rm d, correct}$) \\ 
 Number of steps in test $\taud$ array & 100 \\
 \enddata
%\tablefootnote{We simulated 60 pulse shapes per S/N-$\taud$ pair. }
%\label{table:my_parm}
\end{deluxetable}

\begin{figure*}[t!]
\hspace{-1.2cm}
       \includegraphics[width = 1.1\linewidth]{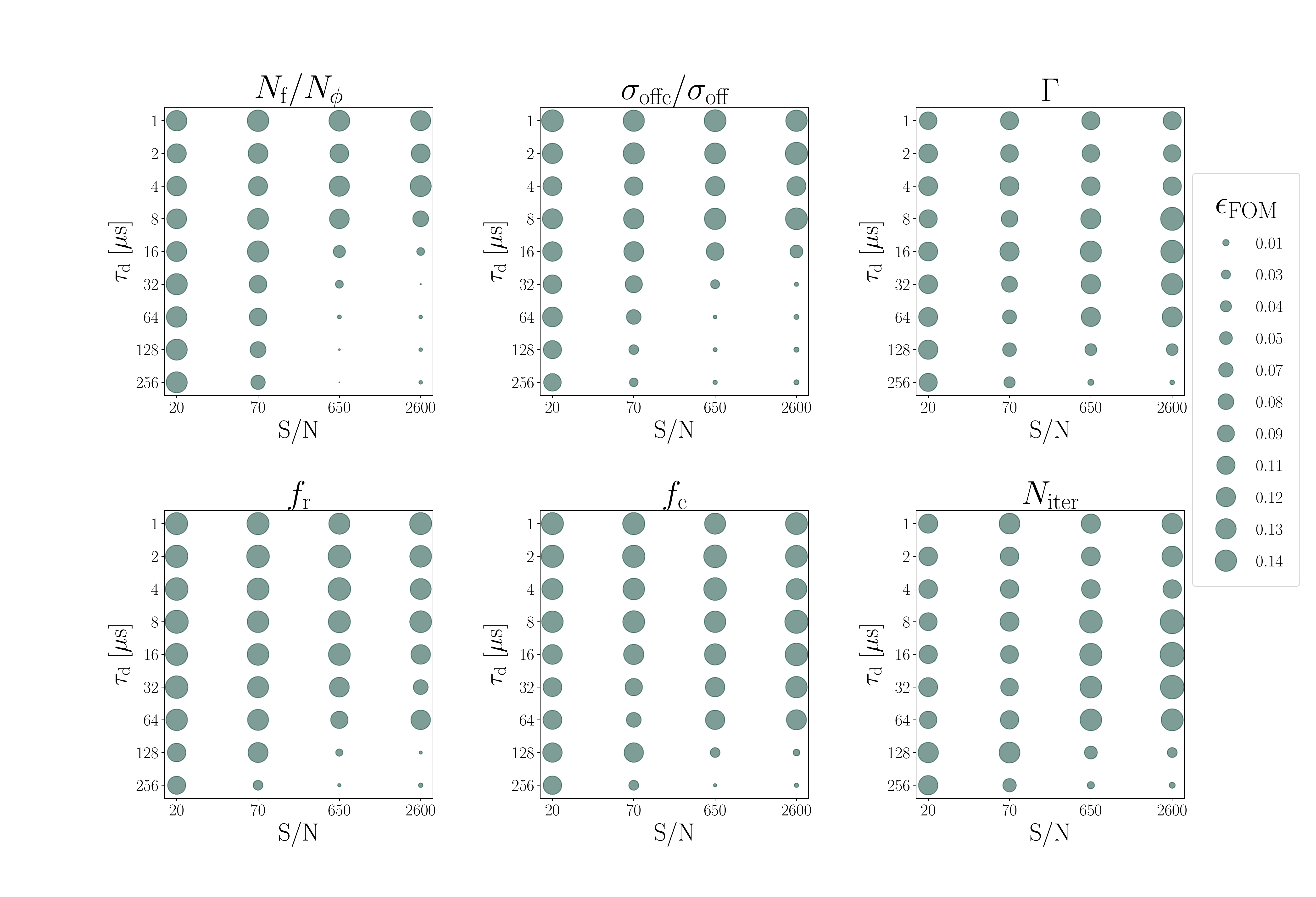}
       \caption{\label{fig:FOM} Overview of the relative performance for each FOM. Smaller circles indicate smaller error bars or better performance of the FOMs on simulated data. These factional averages were computed as described in Equation 13, with $\epsilon_i$ being composed of the $\taud$ returned by only one FOM instead of an unweighted sum of all returned $\taud$ values. These values are labeled as $\epsilon_{\rm FOM}$ in this plot. In general, we see better performance for higher values of $\taud$ and S/N. Interestingly, the $N_{\rm f}$/$N_{\rm \phi}$ and $\sigma_{\rm offc}$/$\sigma_{\rm off}$ FOM appear to perform better than the $f_{\rm r}$ and the $\Gamma$ FOMs which were highlighted in the \citet{Bhat_2003} paper.}
\end{figure*}

%Returning the average error of the returned value of $\taud$ for each $\taud$ and S/N combination instead of the average number of returned $\taud$ values that fall within arbitrarily chosen error bars is more informative about the performance of the algorithm.

%In Figure \ref{fig:dots}, we see a different visualization of the information presented in Figure \ref{fig:error_bar}. Here there are nine subplots, with each plot displaying the recall for each correct $\taud$ across all S/N values. The black error bars represent the average error bar recall shown in Figure \ref{fig:error_bar} in terms of the range of test $\taud$ values iterated over. The green circles indicate the results of each of the 60 simulations for each $\taud$ and S/N pair, with the darker regions indicating areas of higher recall density, indicating the spread of the returned values for each simulation also decreases with higher values of S/N and $\taud$. 

%\begin{figure*}

%       \includegraphics[scale=.35]{test_fig.pdf}
%       \caption{\label{fig:dots} Expansion on the visualization of average error bars of CLEAN recall. The error bar values in Figure \ref{fig:error_bar}  are shown as the black error bars in this figure, with the average error bars from each figure of merit for each of the 60 simulations shown by the green circles. Darker areas show regions of higher density of returned error bars.}
%\end{figure*}

In Figure \ref{fig:FOM}, we show how well each individual FOM performs, with each panel showing the recovery over the full range of S/N-$\taud$ pairs.
%visualize the recall for each figure of merit, with each subplot showing the performance over the full range of $\taud$ and S/N pairs. 
Smaller dots indicate smaller error bars, and thus a more accurate performance. Poor performance from one FOM will impact our averaged recall, as an unweighted average is currently employed. We see that in general, the performance of each FOM improved with higher S/N or $\taud$ like the average, though not all behave equally. For example, the $N_{\rm f}/\Nphi$ and $\sigma_{\rm offc}/\sigma_{\rm off}$ appear to perform better at somewhat lower $\taud$ than the other FOMs. While the skewness $\Gamma$ does not perform as well at the high S/N-$\taud$ end, it does perform marginally better than the previously mentioned two FOMs otherwise. In future works, we will explore developing weights for each FOM in constructing the average $\epsilon_{\rm ave}$ to improve the average accuracy of the algorithm.

\subsection{Testing Secondary Parameter Contributions to Recall Error}

While we assumed the main contributors to the effectiveness of our algorithm to be our primary parameters, $\mathrm{S/N}$ and $\taud $, we wanted to ensure that secondary parameters were not significant contributors to our recall error. We created a small-scale parameterization set via simulation of a base pulse profile with $\taud = 256$ bins and $\mathrm{S/N} = 2600$. We chose very large values for both $\taud$ and S/N as the method was able to reliably recall the correct $\taud$ for large $\taud$ and S/N values (see Section 3.2). This set was used to determine how the number of bins in our observation, the FWHM of the intrinsic pulse, and the user-defined step size and range of the test $\taud$ array affected the algorithm's performance. Additionally, our previous data set assumed an intrinsic pulse with similar parameters to B1937+21 only. Therefore, an additional motivation for probing these secondary parameters was to determine if we can extrapolate our results to observations of other pulsars with varying FWHMs and numbers of phase bins.

For these parameterization runs we used the base values shown in the second column of Table \ref{table:sim_1} and iterated over the values shown in the third column. We run 20 simulations for each variation, which gave insight into these parameters' contribution to our recall and allowed for exploration into the expected larger contributions of $\taud$ and S/N to the recall error.

\begin{deluxetable*}{lll}
\centering 
\tablecolumns{3}
\tablecaption{Secondary Parameters Tested}
\tablehead{
\colhead{Parameter} & \colhead{Base Value} & \colhead{Values}
\label{table:sim_1}
}
\startdata
  Number of phase bins & 256 &  $\left\{128, 256, 512, 1024, 2048\right\}$\\ 
 FWHM & $\frac{1}{8}N_\phi$ & [$\frac{1}{64}$, $\frac{1}{32}$, $\frac{1}{16}, \frac{1}{8}, \frac{1}{4}, \frac{1}{2}$]$N_\phi$ \\
Range of $\taud$ & (0.5 $\tau_{\rm d, correct}$ -- 1.5 $\tau_{\rm d, correct}$) & (0.1 $\tau_{\rm d, correct}$ -- $\tau_{\rm d, correct}$), (0.1 $\tau_{\rm d, correct}$ -- 2.0 $\tau_{\rm d, correct}$),  \\
& & (0.4 $\tau_{\rm d, correct}$ -- 1.6$\tau_{\rm d, correct}$), (0.5 $\tau_{\rm d, correct}$ -- 1.5 $\tau_{\rm d, correct}$) \\
 Number of steps in $\taud$ array & 100 & [10, 20, 50, 100, 200] \\
 \enddata
\end{deluxetable*}

%One exception is for the range of test $\taud$ value runs. Instead of holding the number of steps at 20 for these simulations, we instead hold them to 100 steps to ensure the random chance of choosing the correct $\taud$ to be $\frac{1}{100}$. This is done in an effort to ensure that the only contributing factor in any change in the recall is the range of test $\taud$. 

\begin{figure}[h!]
    \centering
         \includegraphics[width = 1.1\linewidth]{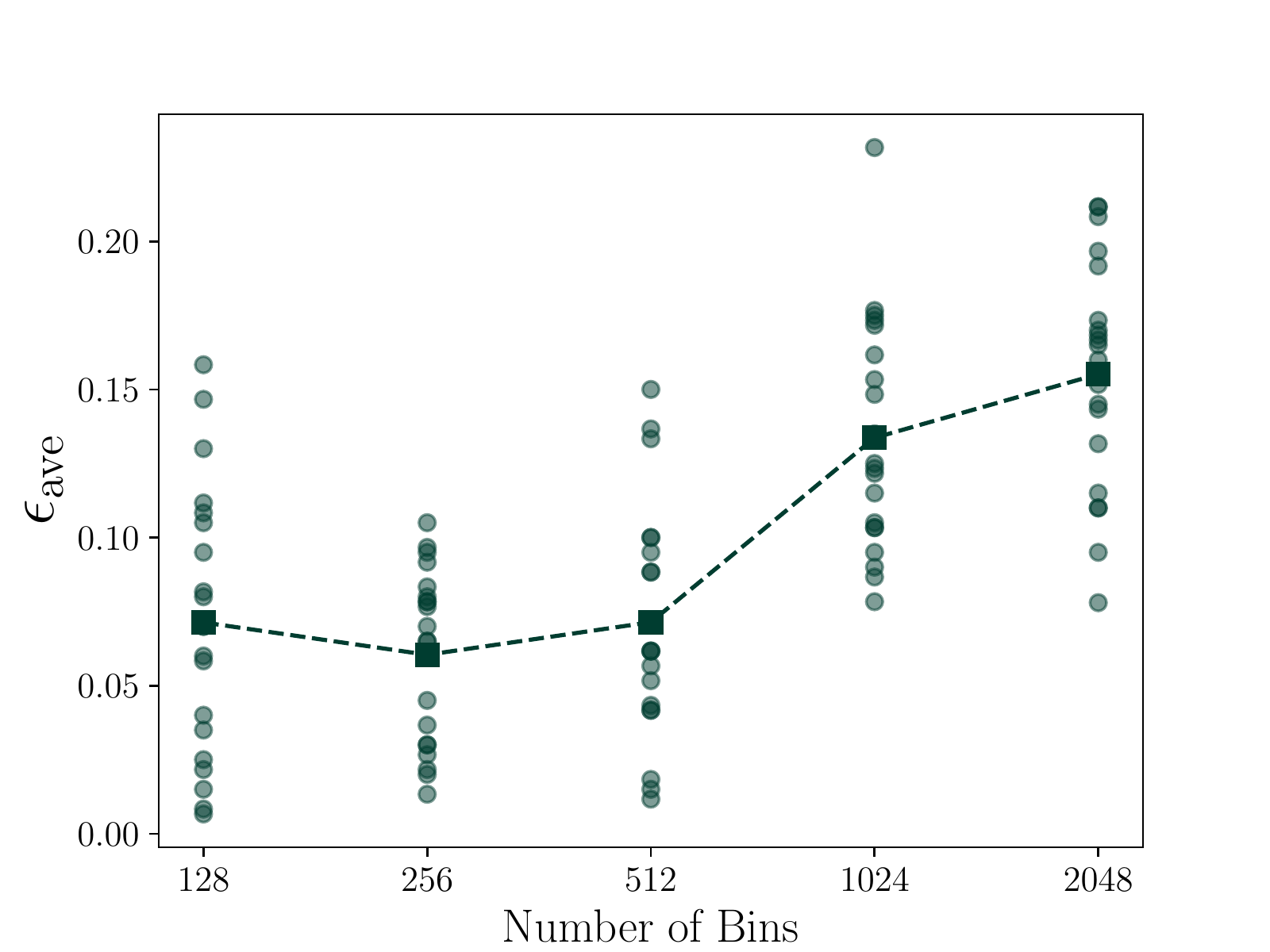}
       \caption{\label{fig:vary_bins} Results of parameterization runs with changing number of phase bins. The y-axis shows the fractional average error bar size across 20 simulations for each bin value. The average recall error is denoted by the fully opaque green squares connected by the dashed line. The lighter circles indicate the recall error from each run. The average error bars range within 10$\%$.}
        %\hspace*{-3.6cm}  
       
  \end{figure}

As many pulse profiles are recorded with a different number of phase bins \citep[see e.g.,][]{EPN}, we tested to see how the phase resolution of the observation affected our recall. We simulated data sets with $\Nphi$ ranging from 128 to 2048. In Figure~\ref{fig:vary_bins}, we see good agreement between the individual recall values for each run and the averages varying within 10\%. Thus, the minor variations in the average recalls could be explained by our limited number of runs resulting in incomplete coverage of the algorithm's performance and we therefore assumed that the number of bins in the observed pulse profile was not a significant contributor to our total recall.

%It is to be noted that most observational data has 1024 bins or higher, with most NANOGrav data using exclusively 2048 bins for their pulsar observations. Therefore, we expect the number of bins used to be fairly consistent across observational data. 

%we see a sharp increase in the average error bar size between 128 to 512 bins, and then a leveling off. This is likely due to the $N = \mathrm{bins}$ cutoff as described in equation 5.16, where the CLEAN algorithm cuts off either when the max of the profile falls below the noise or when the number of CLEAN iterations equals the bin size. Since we are dealing with high S/N data, with a smaller cut-off due to fewer bins we are less likely to see noise components being treated as part of the on-pulse region and thus produce artifacts in the final result. With this being said, especially in NANOGrav data, it is common for data to be binned into 512 or greater bins, as the number of bins is equivalent to the resolution of observational data. In this range, the recall is only affected by less than 2 percent between 512 and 2048 bins, thus ensuring that the results for data binned at 512 versus 2048 bins will likely be relatively unaffected by this choice.  

   \begin{figure}
      \includegraphics[width = 1.1\linewidth]{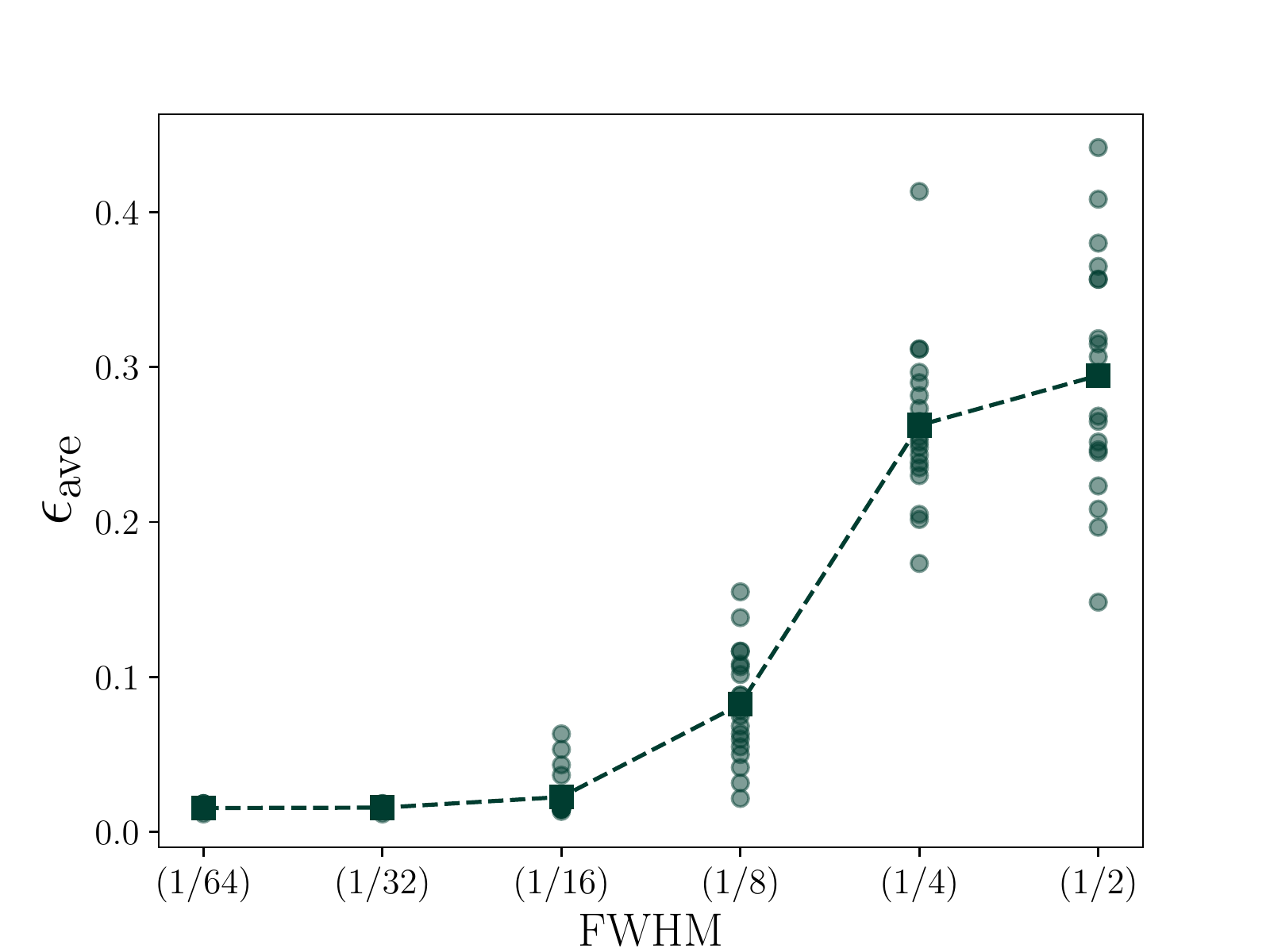}
       \caption{\label{fig:parm_FWHM} Results of parameterization runs with changing values of the FWHM of the intrinsic pulse, as a fraction of the observation length. We see for FWHMs less than $1/4$, the fractional average error bar changes less than 10\%, with the $\epsilon_{\rm ave}$ change still under 30\% for the FWHM being $1/2$ the observation window. This effect has been seen with the CS approach as well.}

\end{figure}

Results of testing how the FWHM of the intrinsic pulse affected our recall are shown in Figure~\ref{fig:parm_FWHM}, and reveal a large, though not unexpected, range in $\epsilon_{\rm ave}$ between the FWHMs tested. As the FWHM increases, the pulse takes up an increasing fraction of the observation window, thus making the CLEAN cutoff criterion of falling below the off-pulse noise level less effective. This result is corroborated by the findings of \cite{Jones_2013} when they found the CS method to be less effective on wider pulses.

%Our next test was to determine if the FWHM of the intrinsic pulse affects our recall. As we can see in Figure \ref{fig:parm_FWHM}, although there is around a 30 percent change in the average recall error bars between the FWHM being $\frac{1}{64}$ of the period of the pulsar (in this case the length of our observation) and $\frac{1}{2}$ of the period of the pulsar, this is not entirely unexpected. When the FWHM of the intrinsic pulse is $\frac{1}{2}$ the period, the pulse takes up the entire window of the observation, thus identifying the off-pulse region of the pulse is impossible, resulting in our CLEAN cutoff criterion of falling below the off-pulse noise level becoming less effective. Thus, this result is not surprising and is corroborated by the findings of \cite{Jones_2013} when they found cyclic spectroscopy to be less effective on wider pulses. Therefore, we can assume that while the results of simulating a single pulsar may not be able to be extrapolated to all pulsars, especially those with very wide pulses, there are still many pulsars that the simulated results should hold for, with the average recall varying by less than 10$\%$ for FWHM values between $\frac{1}{64}$ and $\frac{1}{8}$ of the pulsar period.  

\begin{figure}[h!]
    \centering
     \includegraphics[width = 1.1\linewidth]{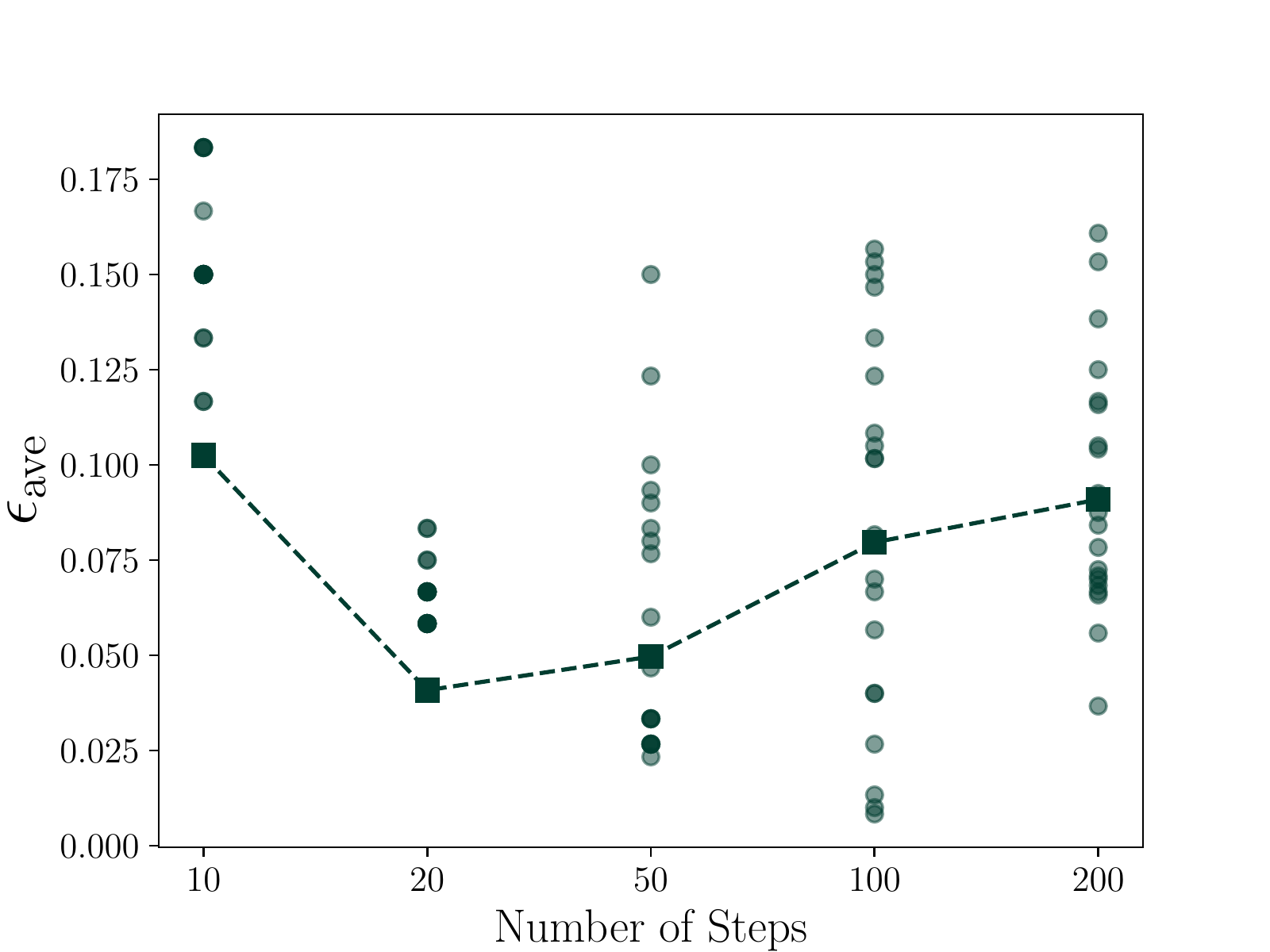}
       \caption{\label{fig:parm_steps} Results of parameterization runs with changing numbers of steps in the test $\taud$ ranges. Although there is some range in the average errors returned, all values agree within $\approx$5\%.}
    
        %\hspace*{-3.6cm}  
      
    \end{figure}

In Figure \ref{fig:parm_steps}, we see the contribution of the number of steps or interchangeably the step size of the test $\taud$ array to our recall error. We included in this analysis a correction factor of $\Delta \taud/2$, the largest base error induced due to large step sizes resulting in the correct $\taud$ not being directly tested. For example, if the correct $\taud$ is 10.5 $\mu\mathrm{s}$, and our test $\taud$ array only samples every $\Delta \taud = 1 \mu\mathrm{s}$, an error of 0.5 $\mu\mathrm{s}$ will be introduced, thus, we added this factor of $\Delta \taud/2$ to our $\epsilon_{\rm ave}$ to more conservatively estimate our uncertainties. The fractional average error bars returned vary within 5\%, therefore we concluded that the number of steps in the test $\taud$ array was not a large contributor to the overall recall error.

%Next, we investigate how the number of steps, or in this case also the step size, of the test $\taud$ array influences our recall. Here, we must take into account the error induced by having a very large step-size, resulting in the correct $\taud$ value not being directly tested which is accounted for be modifying the returned values by $\frac{- \Delta \tau}{2}$, the largest base error induced due to the correct $\taud$ being halfway between two of the $\taud$ values being tested. In Figure \ref{fig:parm_steps}, we see the results of these runs. The average error bars returned vary within 5$\%$. We note a very similar range of returned values for each individual run. Overall however, the range of returned error bars is quite small, and we can therefore conclude that the number of steps in the test $\taud$ array is not a large contributor to the overall recall error. 

%While the recall is much worse for the higher number of steps, we can be sure at this level that random chance is not a larger factor than our algorithms working properly. Thus, larger numbers of steps in our test $\taud$ array should ideally be used for our larger simulation work to ensure that the random chance of choosing the correct $\taud$ is not a larger contributing factor than the actual performance of the algorithm. 
\hfill
   \begin{figure}
          \includegraphics[width = 1.1\linewidth]{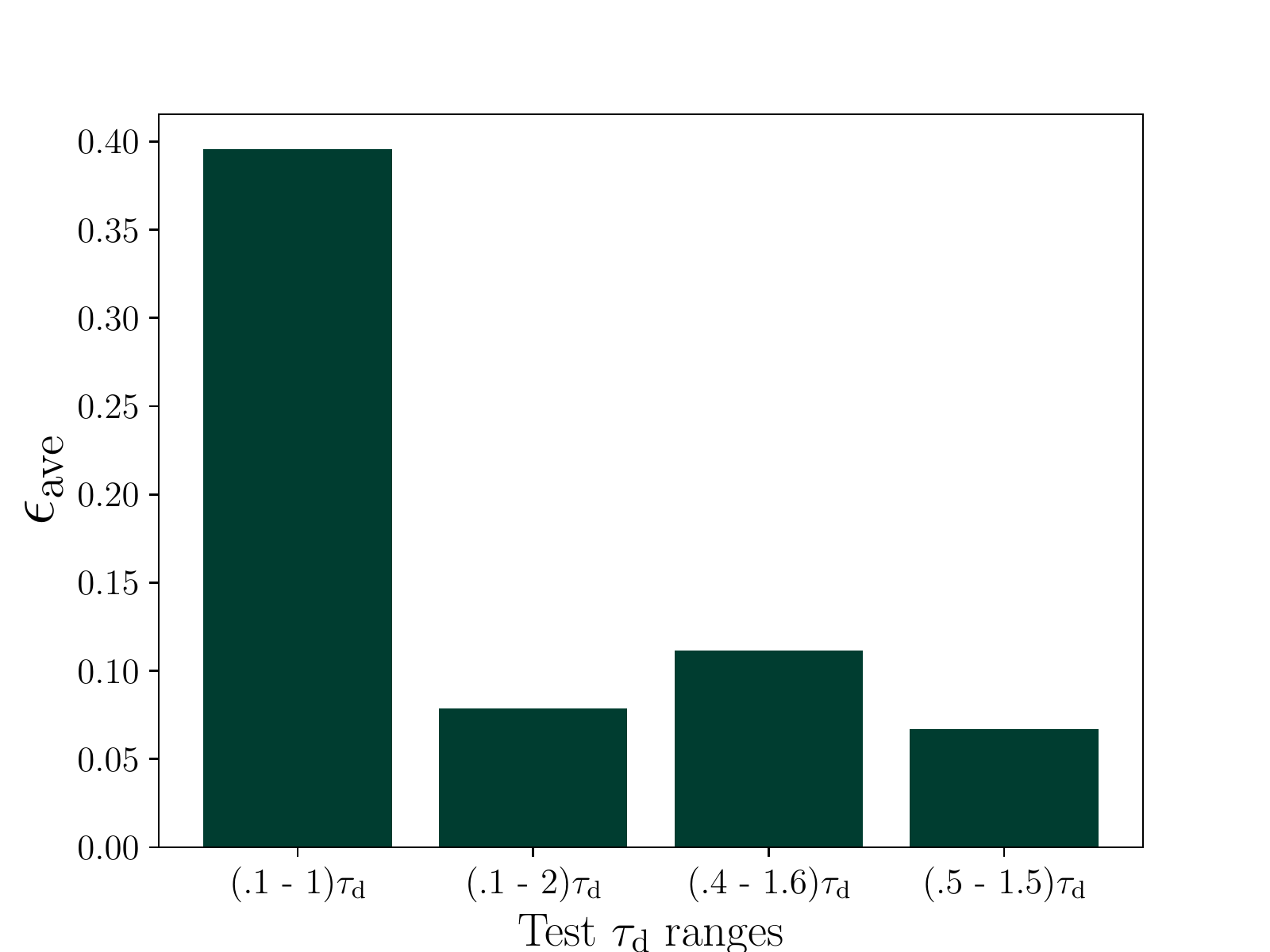}
       \caption{\label{fig:ranges} Results of parameterization runs with changing ranges of test $\taud$ iterated over. We see that as long as the correct $\taud$ is included within the range of test $\taud$ iterated over, the range of test $\taud$ was not a large contributor to the overall recall error.}
       
\end{figure}

Finally, we parameterized the contribution of the range of test $\taud$ iterated over to our recall. In Figure \ref{fig:ranges}, we see that ranges that barely included the correct $\taud$ (second bar) result in poor performance as expected. This results from the shapes of the FOMs not being fully covered over the correct injected $\taud$. Other ranges that include the correct $\taud$ have recalls within about 5\% of each other, even when the ranges iterated over are much larger. Therefore, while more computationally intensive, we recommend running CLEAN over a large range of $\taud$ values to ensure the best estimate is chosen.

%that our choice for $\taud$ is based on not being fully expressed, and ranges that include $\taud$ having recalls within about 5$\%$ of each other, even when the ranges iterated over are much larger.

With the results of these runs, we see that these secondary effects have some small impact at large S/N and $\taud$, but otherwise the most prominent influences on the recall of CLEAN deconvolution are the S/N and $\taud$ of the data. While there are some variations in the average recall for each of the parameters we tested, the average recalls varied within 10\% or less for most tests, with the notable exceptions: large FWHMs and test $\taud$ range that barely included the correct value of $\taud$, both expected. We can also conclude that the results using one simulated pulsar can be extrapolated to both simulations of other pulsars and real observational data.

%We will now delve into how these two factors affect the recall of the CLEAN deconvolution algorithm and compare our findings with those presented in \cite{Dolch_2021}. 

\section{Applying CLEAN to PSR J1903+0327}\label{sec:form}

\label{sec:J1903}

%\textbf{This part is bad rn, I'm working on explaining the plots. }

%From our parameterization, we know that our CLEAN deconvolution requires pulse profiles with high S/N and $\taud$ values, as well as favoring symmetric intrinsic pulses.

%The average S/N of data publicly available on the European Pulsar Network \cite{EPN} is less than 10, and we have determined that CLEAN starts to become effective at much higher S/N levels. While smoothing techniques such as applying a filter could be used, this still would not bring the majority of the publicly available data into the S/N range needed. 

To demonstrate the efficacy of our algorithm, we tested CLEAN on real data from the pulsar J1903+0327. PSR J1903+0327 is a millisecond pulsar that has been monitored by pulsar timing array collaborations such as the North American Nanohertz Observatory for Gravitational Waves (NANOGrav; \citealt{Arzoumanian_2021}) in the effort to detect low-frequency gravitational waves. While these collaborations self-select for pulsars with low amounts of pulse broadening (narrower pulses have higher timing precision), PSR J1903+0327 has some of the most prominent scattering in these data sets, with the broadening tail visible by eye. With over a decade of timing data on this pulsar, we analyzed the lowest-radio-frequency pulses in the NANOGrav 12.5-year data set over time, where broadening is the strongest, to investigate if variations in $\taud$ are detectable by our algorithm. 

We created six summed profiles on which to deploy our CLEAN algorithm, with one profile corresponding to each year from 2012-2017 in our data set. We restricted the frequency band for each observation to 10 MHz centered at 1200 MHz to mitigate additional broadening if frequency-averaging the pulses together to boost S/N. Each summed profile consists of twelve monthly observations summed via cross-correlation. Cross-correlation is used to ensure the peaks of our profiles are properly aligned in time before they are summed, resulting in the highest possible S/N of the summed profile. This process was performed iteratively, with each new profile being cross-correlated with and then added to the summed profile. 

\begin{figure}[h!]

       \includegraphics[width = 1.1\linewidth]{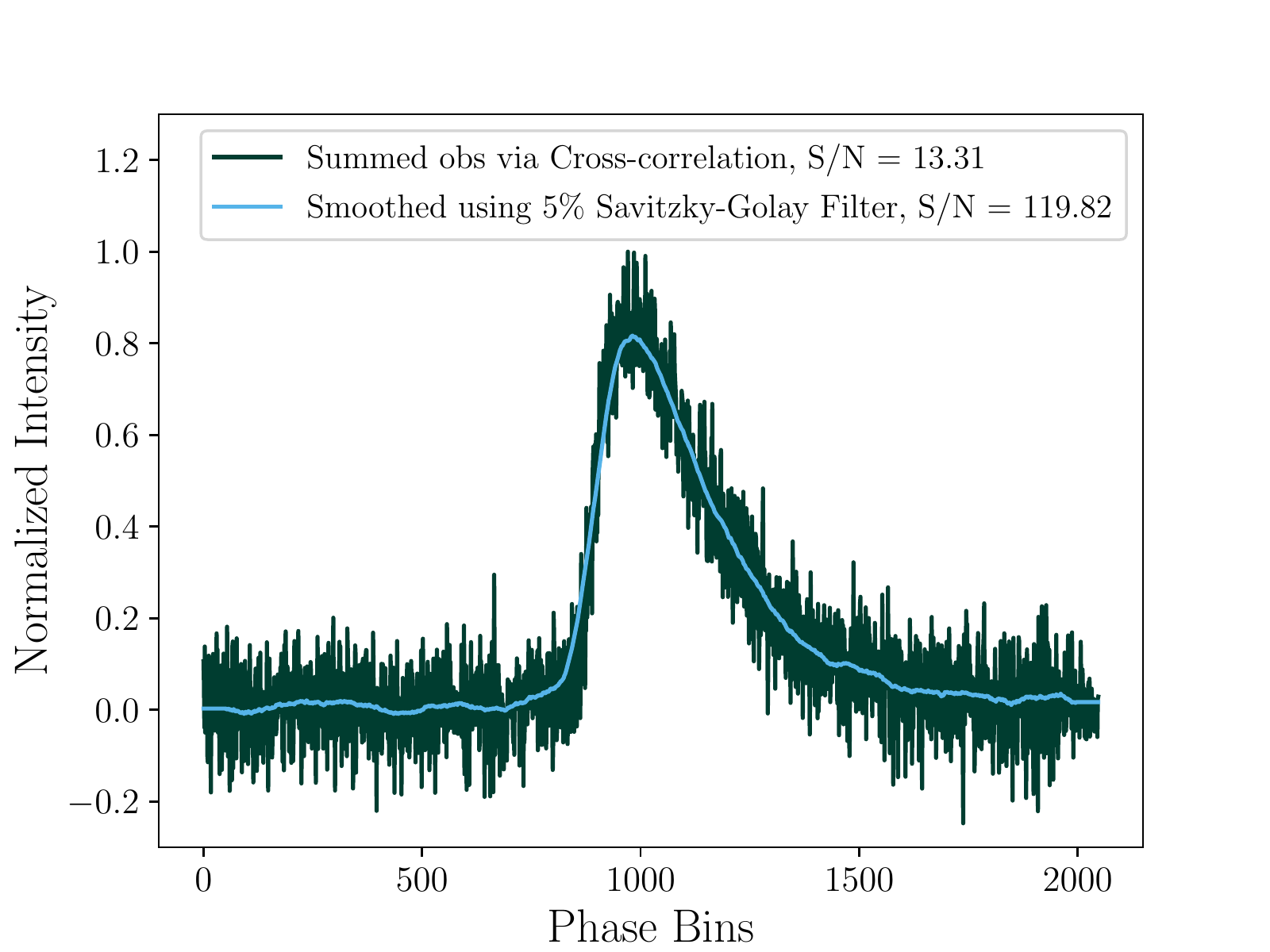}
       \caption{\label{fig:profile} Summed and smoothed pulse profile for J1903+0327 at 1200 MHz from NANOGrav observations during 2014.}
\end{figure}

The refractive timescale of PSR~J1903+0327 is estimated to be between 1 and 2 years \citep{memo8}. Therefore, summing across one year of observations is consistent with the PBF remaining unchanged across this time span. To further increase the S/N values for each profile, we used different Savitzky-Golay filters to smooth the resulting summed profile to the desired S/N level. In Figure~\ref{fig:profile}, we see an example of this summed and smoothed pulse profile.

%At 1 GHz, the broadening tail is estimated to be 0.55 ms. As J1903+0237 has a period of just 2.2 ms, this makes the scattering tail almost a third of the period of the pulsar \cite{ATNF}.

\begin{figure}

       \includegraphics[width = 1.1\linewidth]{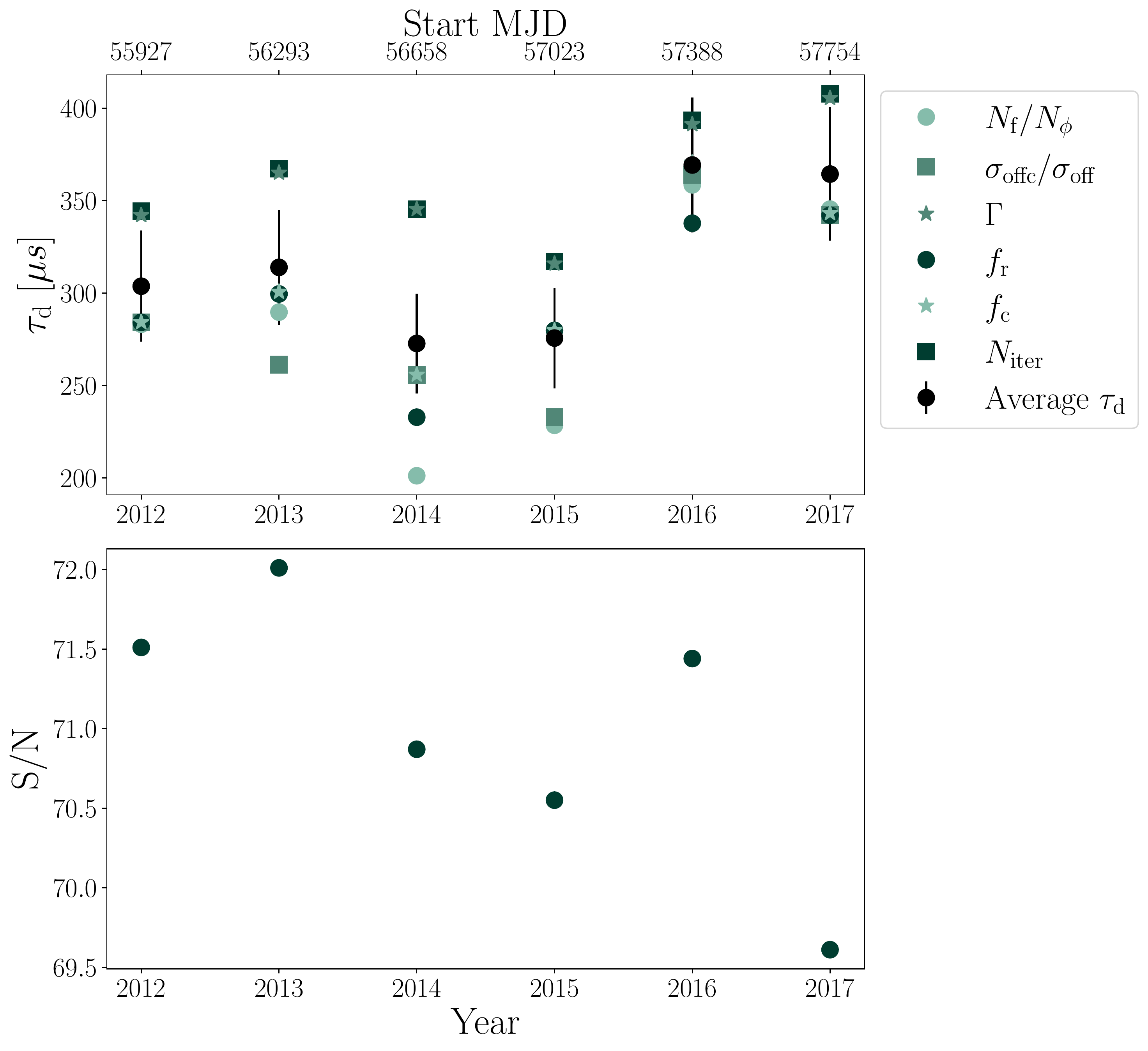}
       \caption{\label{fig:J19_70} Time series for PSR~J1903+0327 from 2012 to 2017 using NANOGrav data. This time series is constructed using a Savitzky-Golay filter with a window filter size chosen to obtain a S/N of around 70. }
\end{figure}

\begin{figure}
       \includegraphics[width = 1.1\linewidth]{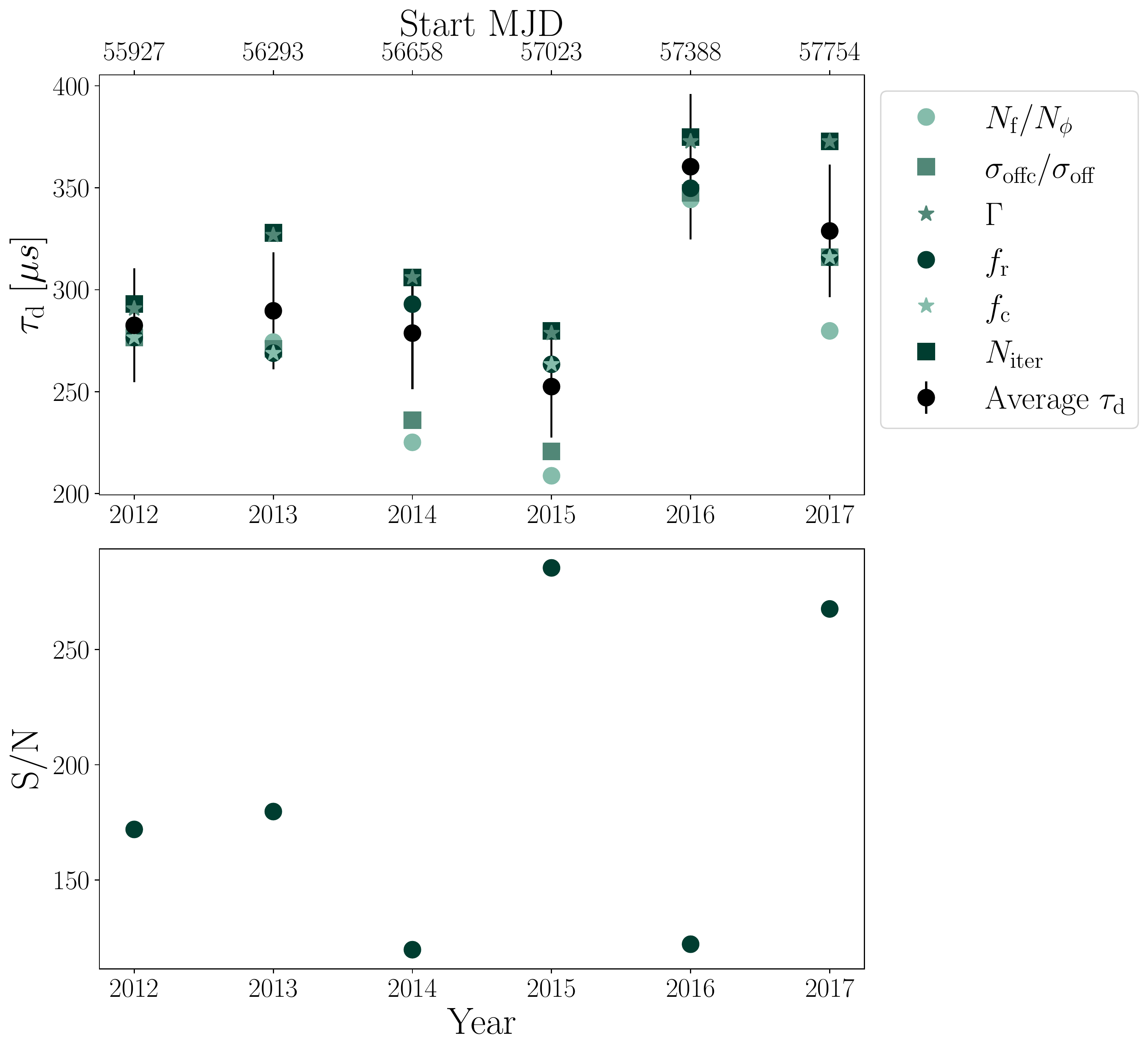}
       \caption{\label{fig:J19_higher} Time series for PSR~J1903+0327 from 2012 to 2017 using NANOGrav data. This time series is constructed using a Savitzky-Golay filter with a window filter size of 5\% of the length of the observation.}
\end{figure}

For our time series analysis, we used two different filtering techniques, both employing a Savitzky-Golay filter using a polynomial of order zero to fit the samples: using a filter window size necessary to achieve S/N of 70 and using a filter window of 5$\%$ of the observation length to achieve a higher S/N. We chose to create time series at two different levels of S/N to showcase the dependence of the algorithm's performance on S/N.  We iterated through test $\taud$ values ranging from 100 to 500 bins for each run, with a step size of one bin. We see the results of these runs in Figures \ref{fig:J19_70} and \ref{fig:J19_higher}, where we converted our returned $\taud$ values into units of microseconds.

%\begin{table}
%\begin{center}
%\begin{tabular}{||c| c ||} 
% \hline
% \textbf{Low S/N FOM} & \textbf{(2012, 2013, 2014, 2015, 2016, 2017)}  \\ [0.3ex] 
% \hline\hline
%  Number of Points & (283.08, 289.64, 201.11, 228.43, 358.50, 345.38)  \\
% \hline
%  RMS & (284.18, 261.22, 255.76, 232.80, 363.96, 342.10)  \\
% \hline
%  Skewness & (342.10, 365.06, 345.38, 315.87, 391.29, 405.50) \\
% \hline
%  Positivity & (284.18, 299.48, 232.80, 279.80, 337.73, 342.10)   \\ 
% \hline
%  Combined & (284.18, 300.57, 255.76, 279.80, 370.52, 343.20) \\ 
% \hline
%  Loops & (344.29, 367.24, 345.38, 316.97, 393.48, 407.68) \\ 
% \hline
%  Average $\taud$ & (303.67, 313.87, 272.70, 275.61, 369.25, 364.33)\\
% \hline  
% 
% \hline\hline
% 
% \textbf{High S/N FOM}  & \textbf{ (2012, 2013, 2014, 2015, 2016, 2017)} \\ [0.3ex] 
% \hline\hline
%  Number of Points  & (281.99, 274.34, 225.15, 208.76, 344.29, 279.80) \\
% \hline
%  RMS & (276.52, 271.06, 236.08, 220.78, 347.57, 315.87) \\
% \hline
%  Skewness  & (290.73, 326.80, 306.04, 278.71, 372.71, 372.71) \\
% \hline
%  Positivity  &(276.52, 268.87, 292.92, 263.41, 349.76, 315.87)   \\ 
% \hline
%  Combined & (276.52, 268.87, 306.04, 263.41, 372.71, 315.87)\\ 
% \hline
%  Loops  & (292.92, 327.9, 306.04, 279.80, 374.89, 372.71)\\ 
% \hline
%  Average $\taud$  & (282.54, 289.64, 278.71, 252.48, 360.32, 328.81)\\
% \hline  

%\end{tabular}
%\caption{}
%\label{table:my_parm}
%\end{center}
%\end{table}

\begin{figure}[t!]
   
        %\hspace*{-3.6cm}
        %\begin{minipage}{0.4\textwidth}
        \includegraphics[width = 1.1\linewidth]{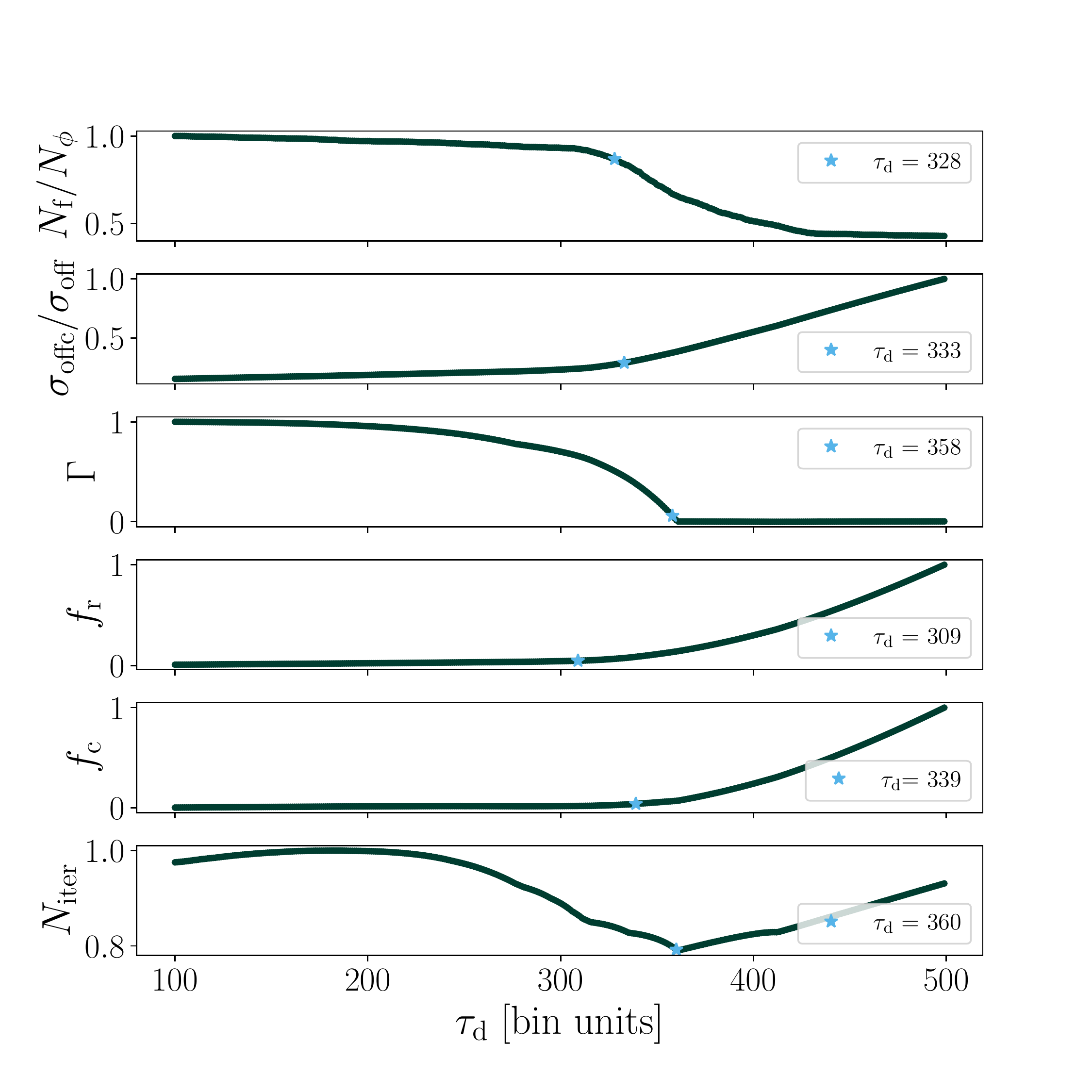}
       \caption{\label{fig:low_SN} Normalized FOMs for PSR J1903+0327 at 1200 MHz from 2016 at S/N = 70. We see a tight grouping for the returned $\taud$ values for each FOM, with a mean value of $\taud = 360.3 ~\mu$s and an error of 10\% based on our simulation runs.
       }
       %and the shape of each FOM makes confirming these returned $\taud$ values by eye possible.  }
    \end{figure}

\begin{figure}
        \includegraphics[width = 1.1\linewidth]{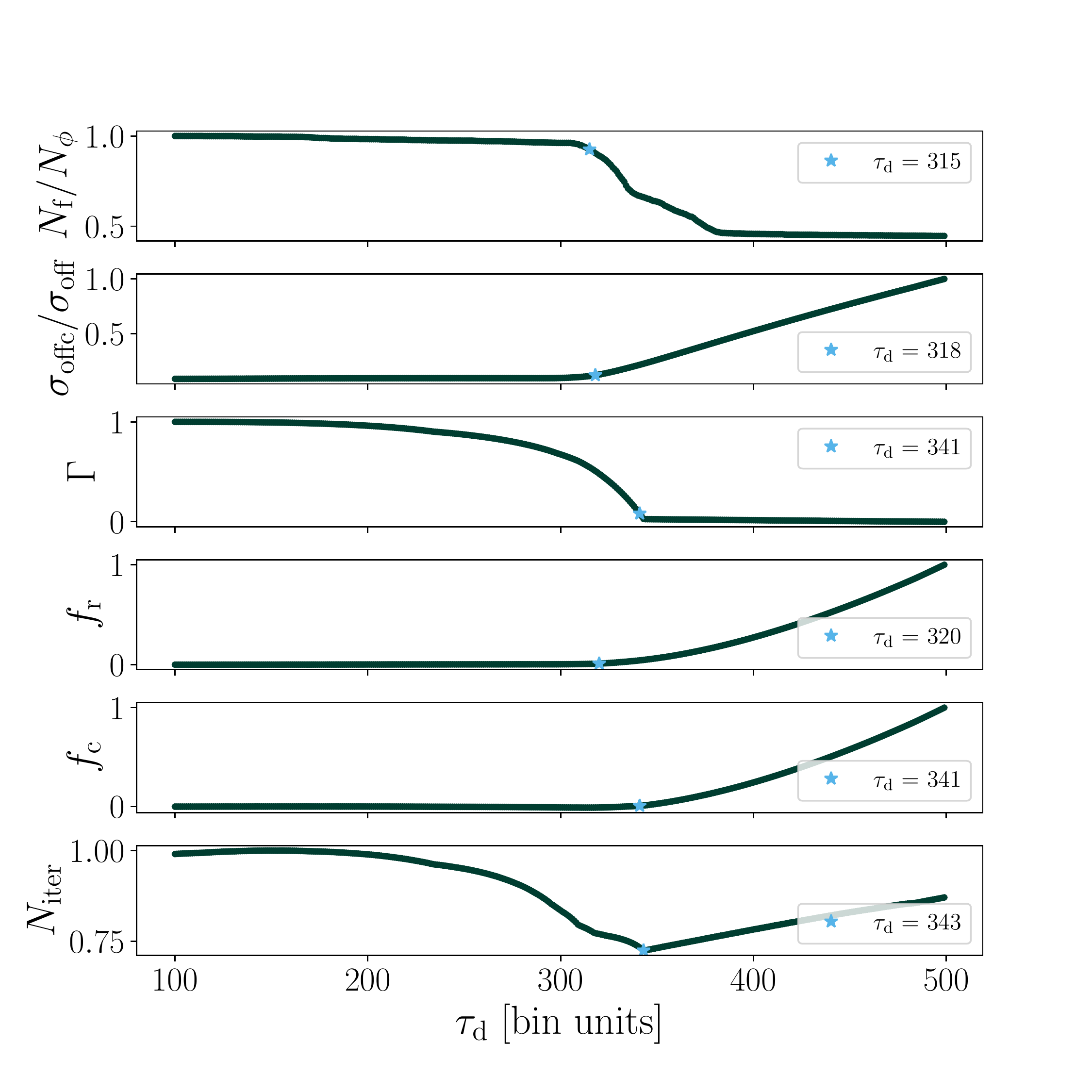}
      \caption{\label{fig:high_SN} Normalized FOMs for PSR J1903+0327 at 1200 MHz from 2016 for high S/N values achieved by using a Savitzky-Golay filter with a window size of 5\% of the length of the observation. We see an even tighter grouping for the returned $\taud$ values for each FOM than compared to the lower S/N FOM, with a mean value of $\taud = 369.3 ~\mu$s and an error between 2\% and 10\% based on our simulation runs. }%Additionally, the shapes of each FOM give even greater confidence in the $\taud$ returned, when compared to those of the lower S/N runs, as there are more distinct visible points of change.   }

\end{figure}

We see what is expected in the FOMs for these time series: greater precision and more visible points of change for the high S/N FOMs. Looking at Figures \ref{fig:low_SN} and \ref{fig:high_SN}, we see examples of how higher S/N results in sharper points of change in the FOMs, thus making choosing the correct $\taud$ a more precise process. While this is true for all FOMs prevented here, this difference can be seen most explicitly in the $\sigma_{\rm offc}$/$\sigma_{\rm off}$ FOM (panel 2), where there is a noticeable location where the slope begins increasing in our larger S/N FOMs, versus our lower S/N FOMs where there is a more gradual increase in the slope of the $\sigma_{\rm offc}$/$\sigma_{\rm off}$ FOM, making the correct $\taud$ more difficult to pinpoint. This increased sharpness of the points of change of our FOMs translated into greater accuracy and better agreement across our FOMs, which can be seen reflected in the tighter clusters around the average returned $\taud$ values in our time series.

%we see the difference between the $\taud$ estimates for each FOM and the overall average of the six.

%we can see the visual difference between the FOM which makes choosing the best value of $\taud$ a more precise process.This can be seen explicitly in the RMS FOM (panel 2), where there is a noticeable location where the slope begins increasing in our larger SN FOMs, versus our lower S/N FOM where there is a more gradual increase in the slope of the RMS FOM, making the correct $\taud$ more difficult to pinpoint.

\hfill

%\begin{figure*}
%    \centering
%   
        %\hspace*{-3.6cm}
%        \begin{minipage}{0.4\textwidth}
%        \includegraphics[scale=0.55]{J1903+0327_1200_MHz_2016_large_range_low_SN.pdf}
%       \caption{\label{fig:low_SN} Figure of Merits for PSR J1903+0327 at 1200 MHz from 2016 at S/N = 70. We see a tight grouping for the returned $\taud$ values for each FOM, and the shape of each FOM makes confirming these returned $\taud$ values by eye possible.  }
%    \end{minipage}\hfill
%   \begin{minipage}{0.49\textwidth}
%        \includegraphics[scale=0.55]{J1903+0327_1200_MHz_5_percent_2016_FOM.pdf}
%      \caption{\label{fig:high_SN} Figure of Merits for PSR J1903+0327 at 1200 MHz from 2016 for high S/N values achieved by using a Savitzky-Golay filter with a window size of 5$\%$ of the length of the observation. We see an even tighter grouping for the returned $\taud$ values for each FOM when compared to the lower S/N FOM. Additionally, the shapes of each FOM give even greater confidence in the $\taud$ returned, when compared to those of the lower S/N runs, as there are more distinct visible points of change.   }
 %    \end{minipage}
%\end{figure*}

We also note some interesting results of this time series analysis, particularly the dip in 2015, followed by a drastic increase the following year. However, coupled with the unusual scattering indices measured in \citet{memo8}, it is clear that an exponential PBF is not supported along this line of sight and a more complex model is necessary (Geiger et al. in prep). Nonetheless, we have shown via this analysis that not only does our CLEAN algorithm perform as expected on observational radio pulsar data given our et of assumptions, but also that employment of this algorithm holds potential for scientific insight into the ever-changing ISM.

%It is to be noted that for this analysis, the FOMs were reviewed by eye to determine our pulse broadening timescale. While our third derivative method was deemed sufficient for large-scale parameterization, by eye analysis provides more accurate results, and is recommended for observational data analysis. 

\section{Future work and Conclusions}

\label{sec:future}
Within this work, we discussed our motivations, introduced CLEAN deconvolution as presented in \cite{Bhat_2003}, discussed our results and products of our implementation of CLEAN, our parameterization work, and results on observational data of PSR J1903+0327. Through our parameterization work, we have concluded that our replicated CLEAN algorithm works as expected: the main factors that influence the recall of the algorithm are the S/N and $\taud$ of the pulse profile, and higher values of S/N and $\taud$ resulting in better recall. We have produced an algorithm that we can confidently deploy on larger sets of observational data. To that end, we have presented a brief analysis of PSR J1903+0327 at two S/N levels and discussed our findings, showing that our methods prove to be effective on observational data from radio pulsars and can thus provide insight into the time-dependence on pulse-broadening timescales for many pulsars after automatic deployment.

Moving forward, we aim to further develop our CLEAN algorithm into a broadly applicable tool, focusing on improving upon or removing the need for a number of simplifications used within this methods paper. We will also deploy our algorithm on the data set used in \citep{Bhat_2004}, the followup to the original CLEAN method introductory paper, and on additional large-scale data sets \citep[e.g.,][]{Stovall_2015,Bilous_2020}. Using these data sets, we will use our CLEAN algorithm to provide measurements of $\taud$ across multiple frequencies along many lines of sight. This will give us greater insight into both the composition of the ISM and the intrinsic emission of radio pulsars. 
%such as the Long Wavelength Array pulsar database pulsar profiles
 
Within this work, we have extensively tested our algorithm's performance on simulated pulses broadened using a thin-screen model of the ISM for our PBF. Future work will entail testing the effects of different pulse broadening functions, namely PBFs based on thick and uniform medium ISM models, on the performance of our algorithm. In addition, while our third derivative method for determining the intrinsic $\taud$ from our FOMs works well given high levels of S/N and large $\taud$ values, this may not hold for low $\taud$ values and low levels of S/N as the FOMs are not as smooth. Therefore, we will work on improving our automation efforts via the implementation of machine learning, thus allowing our recall rates to better reflect the performance of the algorithm. 

We have greatly simplified radio pulsar emission by assuming symmetric Gaussian intrinsic pulses. However, perfectly symmetric pulses are uncommon in radio pulsars \citep[e.g.,][]{Bilous_2016}. %Pulsars can also have non-symmetric sub-pulses and display frequency-dependent shape changes \citep{Cordes_1978}, \citep{Arzoumanian_2021}. 
Should the intrinsic pulse be non-symmetric, our $\Gamma$ FOM will either be completely ineffective or lead to incorrect values of $\taud$ being chosen. Therefore, we must further probe the effects of non-symmetry on our FOM, and develop new FOMs that do not rely on assumed symmetry.

We have developed a Python-based CLEAN algorithm that behaves as expected, have parameterized the performance of this rebuilt algorithm on a variety of simulated test data sets, have developed a method to automate the process of choosing the correct $\taud$ from our FOM, and have proved the efficacy of our algorithm on observational data. We have also defined areas of improvement for our algorithm, and look forward to continuing to develop a well-rounded method for probing the ISM using radio pulsars.

\begin{acknowledgments}
OY is supported by the National Science Foundation Graduate Research Fellowship under Grant No. DGE-2139292. We graciously acknowledge support received from NSF AAG award number 2009468, and NSF Physics Frontiers Center award number 2020265, which supports the NANOGrav project. We acknowledge Research Computing at the Rochester Institute of Technology for providing computational resources and support that have contributed to the research results reported in this publication.
\end{acknowledgments}

\bibliography{sample631}{}
\bibliographystyle{aasjournal}



\end{document}